\documentclass[%
reprint,
superscriptaddress,
 amsmath,amssymb,
aps,
prb,
]{revtex4-1}

\usepackage{graphicx}% Include figure files
\usepackage{subcaption}
\usepackage{dcolumn}% Align table columns on decimal point
\usepackage{bm}% bold math
\usepackage{braket}
\usepackage[colorlinks]{hyperref}
\usepackage{color}
\usepackage[usenames,dvipsnames]{xcolor}

\usepackage[noabbrev]{cleveref}

\crefname{equation}{}{}
\crefname{figure}{}{}
\crefname{table}{}{}

\usepackage[colorinlistoftodos]{todonotes}

\usepackage{xcolor}

\usepackage{verbatim}

\newcommand{\wc}{\omega_\mathrm{c}}
\newcommand{\wm}{\omega_\mathrm{m}}
\newcommand{\xzp}{x_\mathrm{zp}}
\newcommand{\phizp}{\Phi_\mathrm{zp}}
\newcommand{\Qzp}{Q_\mathrm{zp}}
\newcommand{\Vg}{V_\mathrm{g}}
\newcommand{\Vi}{V_\mathrm{I}}

\newcommand{\Cg}{C_\mathrm{g}}
\newcommand{\Cgone}{C_{\mathrm{g}1}}
\newcommand{\Cgonezero}{C_{\mathrm{g}10}}

\newcommand{\Cgtwo}{C_{\mathrm{g}2}}

\newcommand{\Cb}{C_\mathrm{B}}

\newcommand{\Cj}{C_\mathrm{J}}
\newcommand{\Cjone}{C_{\mathrm{J}1}}
\newcommand{\Cjtwo}{C_{\mathrm{J}2}}

\newcommand{\Csigmaone}{C_{\Sigma 1}}
\newcommand{\Csigmatwo}{C_{\Sigma 2}}

\newcommand{\Csigmac}{C_{\Sigma \mathrm{c}}}
\newcommand{\Csigmaonetwo}{C_{\Sigma 12}}
\newcommand{\Csigmaonec}{C_{\Sigma 1\mathrm{c}}}
\newcommand{\Csigmatwoc}{C_{\Sigma 2\mathrm{c}}}

\newcommand{\Cc}{C_\mathrm{c}}
\newcommand{\Ctot}{C_\mathrm{tot}}
\newcommand{\Lc}{L_\mathrm{c}}
\newcommand{\Lb}{L_\mathrm{B}}

\newcommand{\phionedot}{\dot{\phi}_1}
\newcommand{\phitwodot}{\dot{\phi}_2}
\newcommand{\phithreedot}{\dot{\phi}_3}
\newcommand{\phifourdot}{\dot{\phi}_4}
\newcommand{\phicdot}{\dot{\phi}_\mathrm{c}}

\newcommand{\phie}{\Phi_\mathrm{E}}
\newcommand{\phic}{\phi_\mathrm{c}}

\newcommand{\Ng}{n_\mathrm{g}}
\newcommand{\Ngzero}{n_{\mathrm{g}0}}
\newcommand{\grp}{g_\mathrm{rp}}
\newcommand{\gck}{g_\mathrm{CK}}
\newcommand{\gm}{g_\mathrm{m}}

\newcommand{\gcm}{g_\mathrm{cm}}
\newcommand{\gcp}{g_\mathrm{cp}}

\newcommand{\Ceff}{C_\mathrm{eff}}

\newcommand{\xm}{\hat{x}_\mathrm{m}}
\newcommand{\xc}{\hat{x}_\mathrm{c}}
\newcommand{\pc}{\hat{p}_\mathrm{c}}
\newcommand{\Ec}{E_\mathrm{C}}
\newcommand{\Ej}{E_\mathrm{J}}
\newcommand{\Ejone}{E_{\mathrm{J}1}}
\newcommand{\Ejtwo}{E_{\mathrm{J}2}}

\newcommand{\Lagr}{\mathcal{L}}
\newcommand{\Hamil}{\mathcal{H}}
\newcommand{\kinen}{\mathcal{T}}
\newcommand{\poten}{\mathcal{U}}

\newcommand{\Hc}{\mathcal{H}_\mathrm{c}}
\newcommand{\Hm}{\mathcal{H}_\mathrm{m}}
\newcommand{\Hcm}{\mathcal{H}_\mathrm{cm}}
\newcommand{\Hjj}{\mathcal{H}_\mathrm{JJ}}
\newcommand{\Hch}{\mathcal{H}_\mathrm{ch}}

\emergencystretch=\maxdimen
\begin{document}

%\preprint{APS/123-QED}

\title{Enhancement of the optomechanical coupling and Kerr nonlinearity using the Josephson Capacitance of Cooper-pair box}% Force line breaks with \\

\author{Mohammad Tasnimul Haque}
\affiliation{Low Temperature Laboratory, Department of Applied Physics, Aalto University, P.O. Box 15100, FI-00076 Espoo, Finland}

\author{Juuso Manninen}
\affiliation{Low Temperature Laboratory, Department of Applied Physics, Aalto University, P.O. Box 15100, FI-00076 Espoo, Finland}

\author{David Vitali}
\affiliation{School of Science and Technology, Physics Division, University of Camerino, I-62032 Camerino (MC), Italy}

\author{Pertti Hakonen}
\affiliation{Low Temperature Laboratory, Department of Applied Physics, Aalto University, P.O. Box 15100, FI-00076 Espoo, Finland}

\affiliation{QTF Centre of Excellence, Department of Applied Physics, Aalto University, P.O. Box 15100, FI-00076 Aalto, Finland}

\date{\today}% It is always \today, today,

\begin{abstract}
We propose a scheme for enhancing the optomechanical coupling between microwave and mechanical resonators by up to seven orders of magnitude to the ultrastrong coupling limit in a circuit optomechanical setting. The tripartite system considered here consists of a Josephson junction Cooper-pair box that mediates the coupling between the microwave cavity and the mechanical resonator. The optomechanical coupling can be modified by tuning the gate charge and the magnetic flux bias of the Cooper-pair box which in turn affect the Josephson capacitance of the Cooper-pair box. We additionally show that with suitable choice of tuning parameters, the optomechanical coupling vanishes and the system exhibits purely a cross-Kerr type of nonlinearity between the cavity and the mechanical resonator. This allows the system to be used for phonon counting.
%\begin{description}
%\end{description}
\end{abstract}

\maketitle

%\tableofcontents

%xxxxxxxxxxxxxxxxxxxxxxxxxxxxxxxxxxxxxxxxxxxxxx
% INTRODUCTION
%xxxxxxxxxxxxxxxxxxxxxxxxxxxxxxxxxxxxxxxxxxxxxx
\section{Introduction}

Cavity optomechanics\cite{bib:Aspelmeyer2014} using superconducting microwave circuits\cite{bib:Armour2002,bib:Regal2008,bib:LaHaye2009,bib:Rochelau2010,bib:Teufel2011} is an emerging platform for studies of macroscopic quantum phenomena. In particular, there is a growing research interest in ultrastrong coupling regime \cite{bib:Nation2016,bib:Neumeier2018a,bib:Neumeier2018b,bib:Settineri2018,bib:Liao2020,bib:Kounalaksi2020}  where the strength of the single-photon optomechanical coupling is comparable to the resonant frequency of the mechanical resonator. Optomechanical coupling arises from the radiation pressure force acting on a mechanical resonator and, in the microwave regime, this radiation pressure coupling is intrinsically weak. The weak coupling can be amplified by applying a strong coherent pump to the cavity which linearizes the interaction between the cavity and the mechanical resonator. In the linear interaction limit, the quantum effects are observable only close to the quantum zero point fluctuations. However, a nonlinearity can be introduced into the system and single photon strong coupling regime can be reached allowing rich quantum physics experiments, e.g., preparing non-classical states of light and mechanical resonator \cite{bib:Ashhab2010,bib:Gu2014,bib:Sanchez2018,bib:Marinkovic2018,bib:Ockeloen2018} for potential quantum information processing applications.

In the superconducting circuit architecture\cite{bib:H.Devoret2004}, several types of configurations have been proposed to add nonlinearities into the system such as a transmon qubit\cite{bib:Abdi2015}, SQUID\cite{bib:Shevchuk2017}, quantum capacitance of a nanotube in Coulomb blockade regime\cite{bib:Blien2020}, and Josephson inductance\cite{bib:Heikkila2014,bib:Rimberg2014} which has also been experimentally realized \cite{bib:Pirkkalainen2015}. We propose to employ ``Josephson Capacitance'' of the Cooper-pair box (CPB) which is dual to the operation of Josephson inductance as the nonlinear element for enhancing the optomechanical coupling. Looking from the gate electrode, a CPB can act as a nonlinear capacitive element known as the Josephson capacitance. This capacitance originates from the curvature of the energy bands of the CPB with respect to gate charge \cite{bib:Averin1985,bib:Averin2003,bib:Sillanpaa2005,bib:Duty2005,bib:Persson2010}. Josephson capacitance has been proposed to be utilized as a very sensitive phase detector\cite{bib:Roschier2005} and a pair breaking radiation detector\cite{bib:Shaw2009}. Here, we consider a tripartite system consisting of a microwave cavity, a CPB, and a mechanically movable capacitance. The coupling between the cavity and the mechanical resonator is mediated by the Josephson capacitance of the CPB and it can be tuned by the charge and flux bias of the CPB. We show that with suitably tuned, realistic experimental parameters, the optomechanical coupling can be enhanced by seven orders of magnitude compared to direct optomechanical coupling without the presence of CPB, reaching the ultrastrong coupling regime.

In addition to boosting of the optomechanical coupling, a cross-Kerr (CK) type nonlinearity between the cavity and the mechanics is also formed amidst other nonlinear terms in the system Hamiltonian. %into the cavity which puts a limit on the enhancement we can obtain from this system.
The CK coupling $\gck \hat{n}_\mathrm{a} \hat{n}_\mathrm{b}$ between two resonators a and b, with number operators $\hat{n}_\mathrm{a}$ and $\hat{n}_\mathrm{b}$, can be used for quantum nondemolition measurements of number of quanta in one of the resonators since it directly affects the resonance frequency of the readout resonator. \cite{bib:Imoto1985} In recent years, CK coupling in optomechanical systems has attracted theoretical interest \cite{bib:Khan2015,bib:Sarala2015,bib:Xiong2016,bib:Chakraborty2017} including a recent scheme to enhance the coupling to the order of the cavity linewidth\cite{bib:Yin2018}. Tunable CK interaction has been experimentally demonstrated for superconducting microwave circuits.\cite{bib:Kounalakis2018}

%xxxxxxxxxxxxxxxxxxxxxxxxxxxxxxxxxxxxxxxxxxxxxx
% DESCRIPTION OF THE SYSTEM
%xxxxxxxxxxxxxxxxxxxxxxxxxxxxxxxxxxxxxxxxxxxxxx
\section{Description of the system}

\begin{figure}[htb]
\centering
\includegraphics[width=1\linewidth]{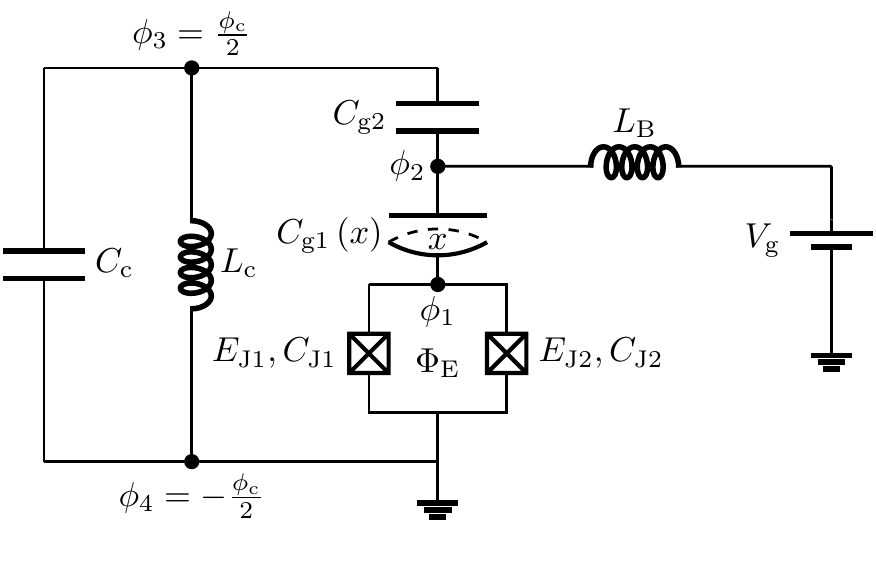}
\caption{Circuit diagram of the investigated optomechanical setup with the gate capacitance of the Cooper-pair box connected in parallel to the microwave cavity capacitance. The elements in the circuit are described in the text.}
\label{fig:ckt}
\end{figure}

A circuit diagram of the setup is presented in Fig. \ref{fig:ckt}. The CPB formed of two Josephson junctions with Josephson energies $\Ejone, \Ejtwo$ and capacitances $\Cjone, \Cjtwo$ couples a mechanical displacement dependent gate capacitance $\Cgone \left(x\right)$ to a superconducting microwave cavity, modeled as a simple LC oscillator in a bias-T configuration. Here, we assume $\Cgone \left(x\right) = \Cgonezero + \frac{\partial \Cgone \left( x \right)}{\partial x}x + \frac{1}{2} \frac{\partial^2 \Cgone \left( x \right)}{\partial x^2}x^2$, where the derivatives are approximated for a parallel plate capacitor.
This configuration allows the CPB to act as a capacitive element and, additionally, blocks the DC gate bias from entering into the microwave cavity and prevents the AC signal leaking out through the DC gate. The inductor $\Lb$ in the bias-T and the capacitor $\Cgtwo$ have impedances such that they do not affect the qubit dynamics. The junction capacitances $\Cjone$ and $\Cjtwo$ and the static part of the gate capacitance give the total single electron charging energy of the qubit, $\Ec = e^2/2 \left( \Cgonezero + \Cjone + \Cjtwo \right)$. Tunneling of a Cooper pair into the qubit island is tunable by the static gate charge given by the number of Cooper pairs in the island $\Ngzero = - \Vg \Cgonezero/(2e)$ where $\Vg$ is the gate voltage. Furthermore, the CPB is split and hence, by applying a magnetic flux $\phie$ to the superconducting loop, the Josephson energy of the qubit can be controlled.

To better illustrate the procedure of deriving the effective Hamiltonian for the system, we assume in the following, without loss of generality, that the Josephson energies of the two junctions are equal, i.e. $\Ejone = \Ejtwo = \Ej/2$, in which case the Hamiltonian of the unperturbed qubit is
\begin{equation}
\label{eq:qubit_Hamiltonian}
    \hat{\mathcal{H}}_\mathrm{QB} = -\frac{1}{2}\sum_{j=1,3} B_j\sigma_j
\end{equation}
with the ground state energy, $E_\mathrm{GS} = -\sqrt{B_1^2 + B_3^2}/2 = -B/2$ where $B_1 = \Ej\cos (\pi\frac{\phie}{\Phi_0})$ and $B_3 = -4\Ec(1-2n_{\mathrm{g}0})$. Here $\sigma_j$ are the Pauli spin matrices and $\Phi_0 = h/2e$ is the flux quantum. In principle, the junction energies can be different, and the explicit full calculations of the system dynamics are presented in the Appendices.

%xxxxxxxxxxxxxxxxxxxxxxxxxxxxxxxxxxxxxxxxxxxxxx
% CIRCUIT MODEL
%xxxxxxxxxxxxxxxxxxxxxxxxxxxxxxxxxxxxxxxxxxxxxx
\section{Circuit model}

Enhancement of the radiation pressure coupling can be understood from the quantum capacitance picture where the effective capacitance of the CPB is affecting the total capacitance of the cavity. For band $k$ of the CPB, the effective capacitance is given by \cite{bib:Duty2005,bib:Sillanpaa2005}
\begin{equation}
\label{eq:effective_capacitance}
    \Ceff^k = \frac{\Cgone \Cj}{\Csigmaone} - \frac{\Cgone^2}{4 e^2}\frac{\partial^2 E_k (\phie,\Ng) }{\partial \Ng^2} ,
\end{equation}
where we denote $\Cj = \Cjone + \Cjtwo$ and $\Csigmaone = \Cgone + \Cj$.

This effective capacitance $\Ceff$, containing both geometric and quantum capacitance contributions of the CPB, is in parallel with the cavity capacitance $\Cc$. As a consistency check, one sees that in the limit of small Josephson energy the effective capacitance approaches the geometric gate capacitance. The resonance frequency of the cavity is then $\wc = 1/\sqrt{\Lc \Ctot}$, where the total capacitance consist of the cavity capacitance in parallel with the second gate capacitance and the effective CPB capacitance, i.e. $\Ctot = \Cc + \left( \frac{1}{\Cgtwo} + \frac{1}{\Ceff} \right)^{-1}$. The radiation pressure coupling is given by the linear expansion of the resonant frequency with respect to the mechanical displacement
\begin{equation}
\label{eq:radiation_pressure_coupling}
    \hbar \grp = -\hbar \frac{\partial \wc}{\partial x}\xzp ,
\end{equation}
where $\xzp$ is the zero point motion of the mechanical displacement.

A straightforward calculation yields
\begin{subequations}
\begin{align}
    \frac{\partial \wc}{\partial x} &= - \frac{1}{2} \frac{\Cgtwo^2}{\left( \Cgtwo + \Ceff \right)^2} \frac{\wc}{\Ctot} \frac{\partial \Ceff}{\partial x} , \label{eq:dwc_dx} \\
    \frac{\partial \Ceff}{\partial x} &= -\frac{\Cj \Cgone \Cgone'}{\left(\Cgone + \Cj\right)^2} + \frac{\Cj \Cgone'}{\Cgone + \Cj} \nonumber \\ 
    &\ \ \ - \frac{\Cgone \Cgone'}{2 e^2} \frac{\partial^2 E_k}{\partial \Ng^2} - \frac{\Cgone^2}{4 e^2} \frac{\partial}{\partial x} \left( \frac{\partial^2 E_k}{\partial \Ng^2} \right) \label{eq:dCeff_dx}
\end{align}
\end{subequations}
with $\frac{\partial}{\partial x} \left( \frac{\partial^2 E_k}{\partial \Ng^2} \right) = \frac{\partial \Ng}{\partial x} \frac{\partial}{\partial \Ng} \left( \frac{\partial^2 E_k}{\partial \Ng^2} \right)$ where $\frac{\partial \Ng}{\partial x} = - \frac{\Cgone'}{2 e} \Vg$.

Plugging Eqs. \eqref{eq:dwc_dx} and \eqref{eq:dCeff_dx} in to Eq. \eqref{eq:radiation_pressure_coupling}, we obtain the final expression for optomechanical coupling
\begin{equation}
\label{eq:grp_circuit}
\begin{split}
        \hbar \grp =& \hbar \xzp \left[ \frac{1}{2} \frac{\Cgtwo^2}{\left( \Cgtwo + \Ceff \right)^2} \frac{\wc}{\Ctot} \right] \\
        &\times \left[ -\frac{\Cj \Cgone \Cgone'}{\left(\Cgone + \Cj\right)^2} + \frac{\Cj \Cgone'}{\Cgone + \Cj} \right. \\
        &\ \ \ \ \ \left. - \frac{\Cgone \Cgone'}{2 e^2} \frac{\partial^2 E_k}{\partial \Ng^2} + \frac{\Cgone^2 \Cgone'}{8 e^3} \Vg \frac{\partial^3 E_k}{\partial \Ng^3} \right].
\end{split}
\end{equation}

\onecolumngrid

\begin{figure}[tbh]
    \centering
    \begin{subfigure}[]{0.45\textwidth}
        \includegraphics[width=\textwidth]{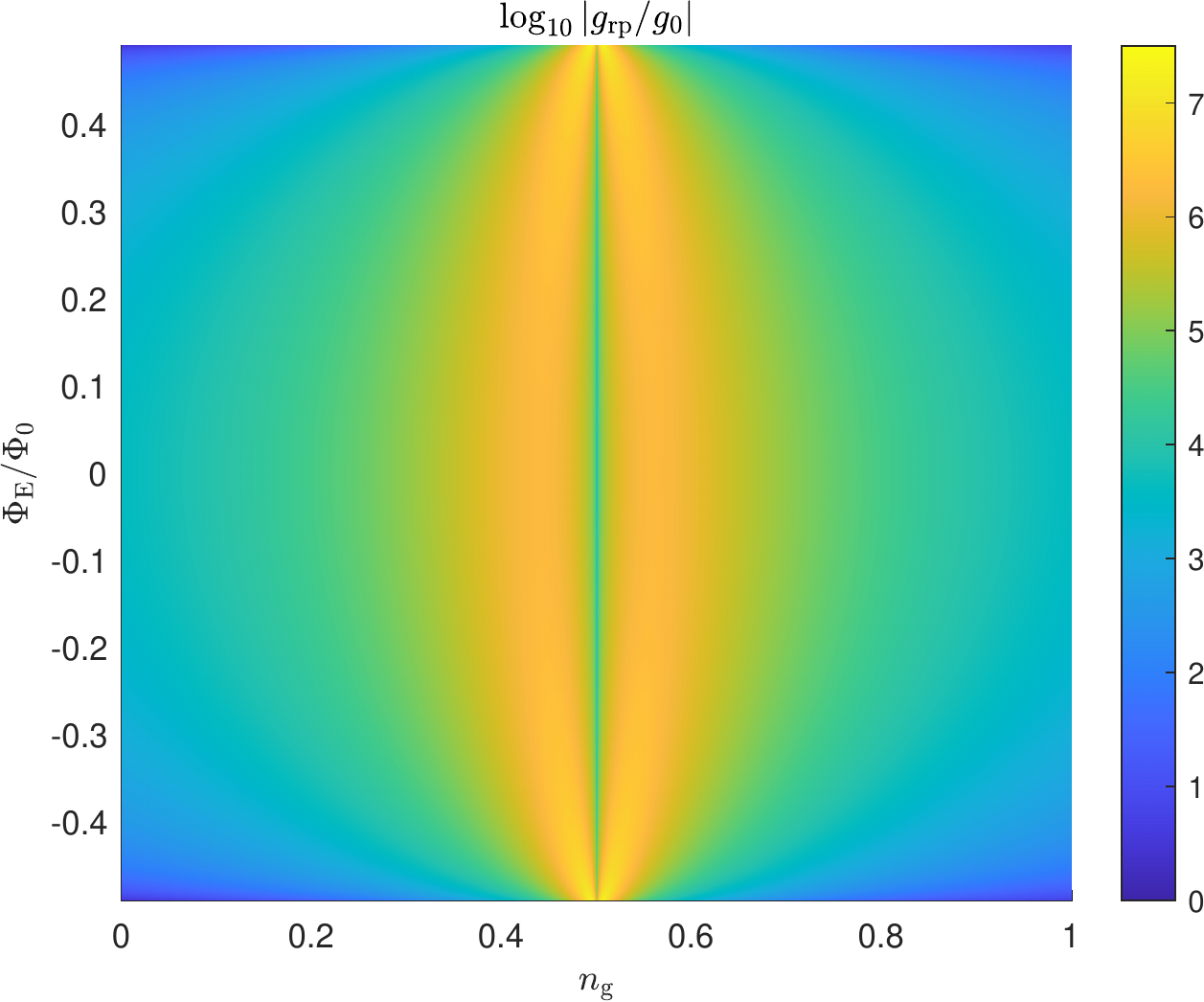}
        \caption{}
        \label{fig:grp_field_plot}
    \end{subfigure}
    \begin{subfigure}[]{0.45\textwidth}
        \includegraphics[width=\textwidth]{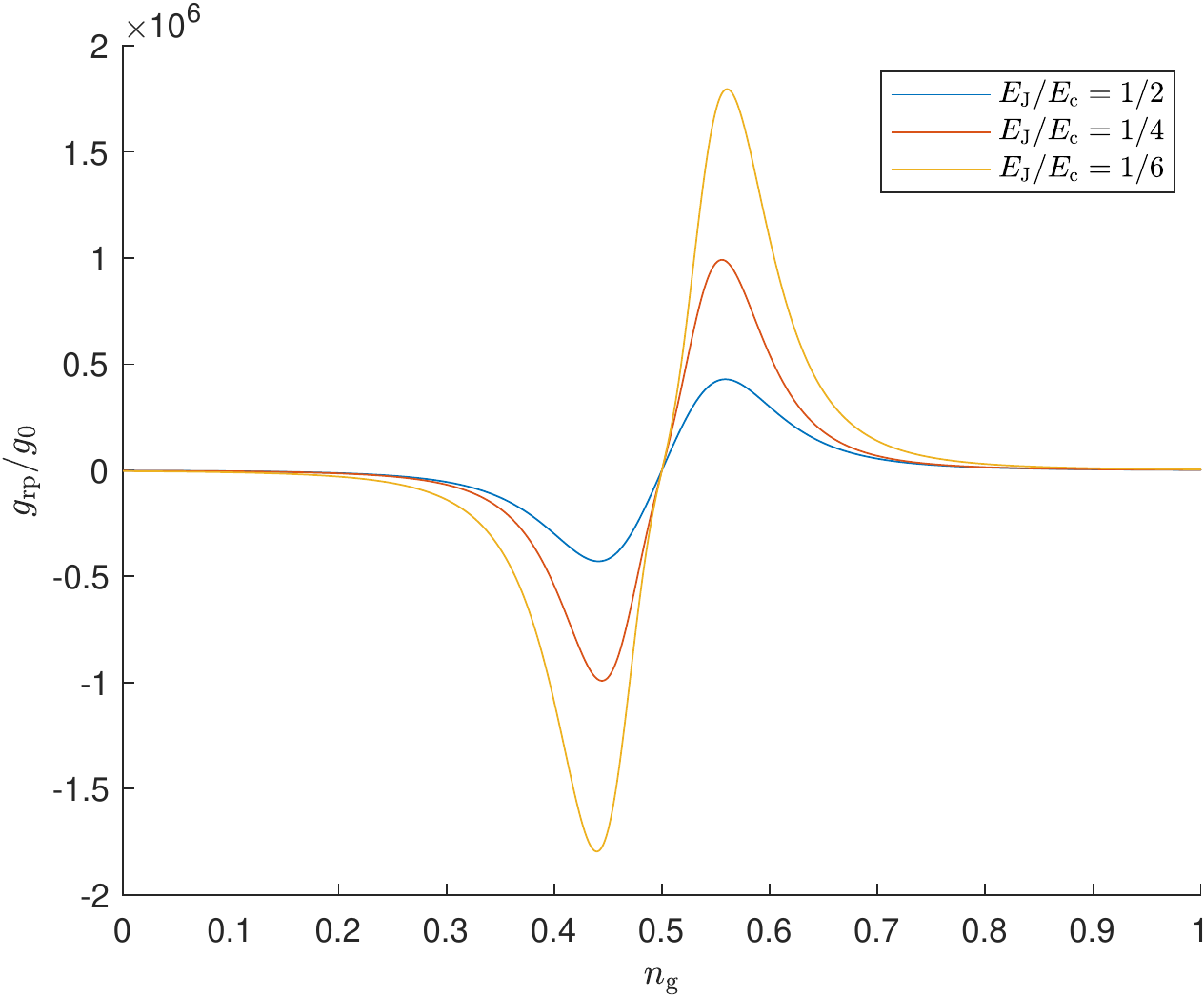}
        \caption{}
        \label{fig:grp_ng}
    \end{subfigure}
    \label{fig:grp_enhanced_full}
    \caption{\textbf{(a)} The enhancement of the optomechanical coupling $\grp$ due to the quantum capacitance compared to direct coupling $g_0$ as a function of flux bias and gate charge. Here $\Ec/h = 30\,$GHz and $\Ej/h = 7.5\,$GHz. \textbf{(b)} The enhancement of the optomechanical coupling for different $\Ej/\Ec$ ratio as a function of gate charge with $\Ej/h = 5\,$GHz at $\phie = 0$.}
\end{figure}

\twocolumngrid

The enhancement of the radiation pressure coupling is compared to the direct optomechanical coupling in the absence of the qubit
\begin{equation}
\label{eq:direct_grp}
    g_0 = -\frac{1}{2} \frac{\Cgtwo^2}{\left( \Cgone + \Cgtwo \right)^2} \frac{\wc}{C_\mathrm{d}} \Cgone' \xzp ,
\end{equation}
where the cavity frequency is influenced by the total capacitance $C_\mathrm{d} = \Cc + \left( \frac{1}{\Cgone} + \frac{1}{\Cgtwo} \right)^{-1}$ in the direct coupling scheme.

For a numerical estimation of the radiation pressure coupling, presented in Fig. \ref{fig:grp_field_plot}, we choose a cavity of resonant frequency $\wc/2\pi = 5\,$GHz with characteristic impedance $Z_0 = 100 \, \Omega$ resulting in capacitance $\Cc = 0.318\,$pF and inductance $\Lc = 3.18\,$nH. The other parameter values chosen are: $\Vg = 10\,$V, $\Ec/h = 30\,$GHz ($\sim 124\,\mu$eV), $\Ej/h = 7.5\,$GHz ($\sim 21\,\mu$eV), thus $\Ej/\Ec = 1/6$. The reason to choose these values for $\Ec$, $\Ej$ is to minimize quasiparticle poisoning which is a well known limiting factor associated with CPB devices. The band energies of CPB are 2$e$ periodic with respect to gate charge modulation. In practice, however, due to  tunnelling of non-equilibrium quasiparticles on and off the CPB island,\cite{bib:Joyez1994,bib:Aumentado2004} the periodicity can change from 2$e$ to 1$e$. To reduce this effect, it is desirable to have $\Ec$ smaller than the superconducting energy gap $\Delta_\mathrm{g}$. Typically qubits are made of superconducting Aluminum (Al) junctions and for Al, critical current ($T_\mathrm{C}$) is $1.2\,$K, the superconducting gap $\Delta_\mathrm{g} \sim 1.76 k_\mathrm{B} T_\mathrm{C} \sim 182\,\mu$eV which is well above the chosen value.

Another crucial factor that influences the coupling strength is the $\Ej/\Ec$ ratio. In Fig. \ref{fig:grp_ng}, we plot the coupling enhancement $\grp/g_0$ against gate charge $\Ng$ for several $\Ej/\Ec$ ratios at flux $\phie=0$ to better illustrate how the enhancement depends on $\Ng$. The maximum coupling is reachable near the charge qubit limit $\Ec \gg \Ej$.

As depicted below in the quantum mechanical treatment of the circuit, Eq.\eqref{eq:sqrt_expansion}, the device exhibits a cross-Kerr type of nonlinearity $\gck a^\dagger a b^\dagger b$. Using the quantum capacitance approach similar to deriving Eq.  \eqref{eq:grp_circuit}, we obtain the cross-Kerr coupling
\begin{equation}
\label{eq:cross-Kerr_coupling}
    \hbar \gck = \hbar \frac{\partial^2 \wc}{\partial x^2} \xzp^2 ,
\end{equation}
where the second order derivative of the cavity frequency is
\begin{equation}
\label{eq:d2wcdx2}
\begin{split}
        \frac{\partial^2 \wc}{\partial x^2} = & \frac{1}{4} \wc \Cgtwo^2 \left( \frac{\partial \Ceff}{\partial x} \right)^2 \\
        & \times \frac{\Cgtwo \left( 4\Cc + 3\Cgtwo \right) + 4\left( \Cc + \Cgtwo \right) \Ceff}{\left( \Cgtwo + \Ceff \right)^2 \left[ \Cc \Cgtwo + \left( \Cc + \Cgtwo \right) \Ceff \right]^2}  \\
        &-\frac{1}{2}\wc \Cgtwo^2 \frac{\partial^2 \Ceff}{\partial x^2} \\
        & \times \frac{1}{\left( \Cgtwo + \Ceff \right) \left[ \Cc \Cgtwo + \left( \Cc + \Cgtwo \right) \Ceff \right]} 
\end{split}
\end{equation}
with

\onecolumngrid

\begin{figure}[tbh]
    \centering
    \begin{subfigure}[]{0.45\textwidth}
        \includegraphics[width=\textwidth]{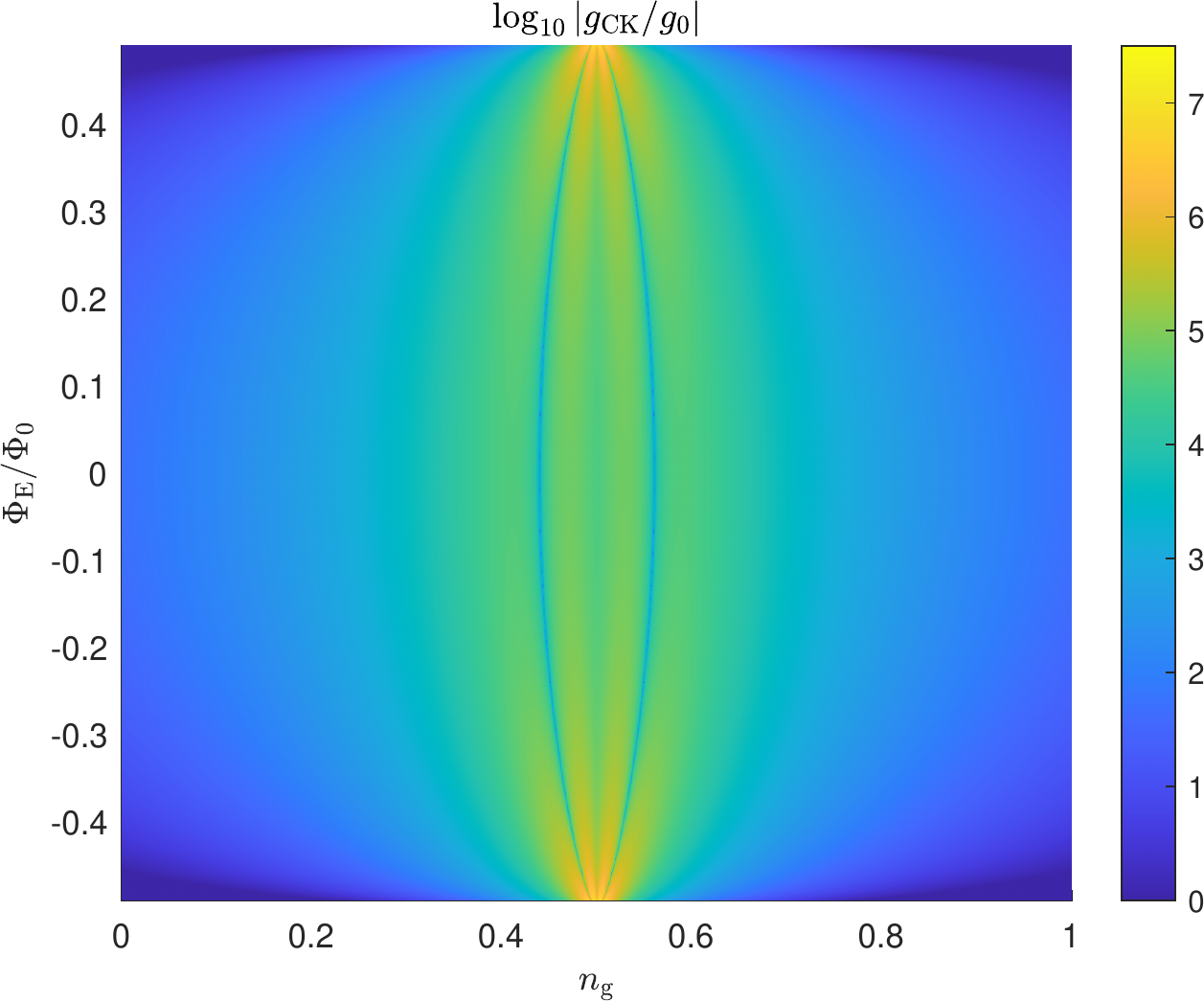}
        \caption{}
        \label{fig:gck_field_plot}
    \end{subfigure}
    \begin{subfigure}[]{0.45\textwidth}
        \includegraphics[width=\textwidth]{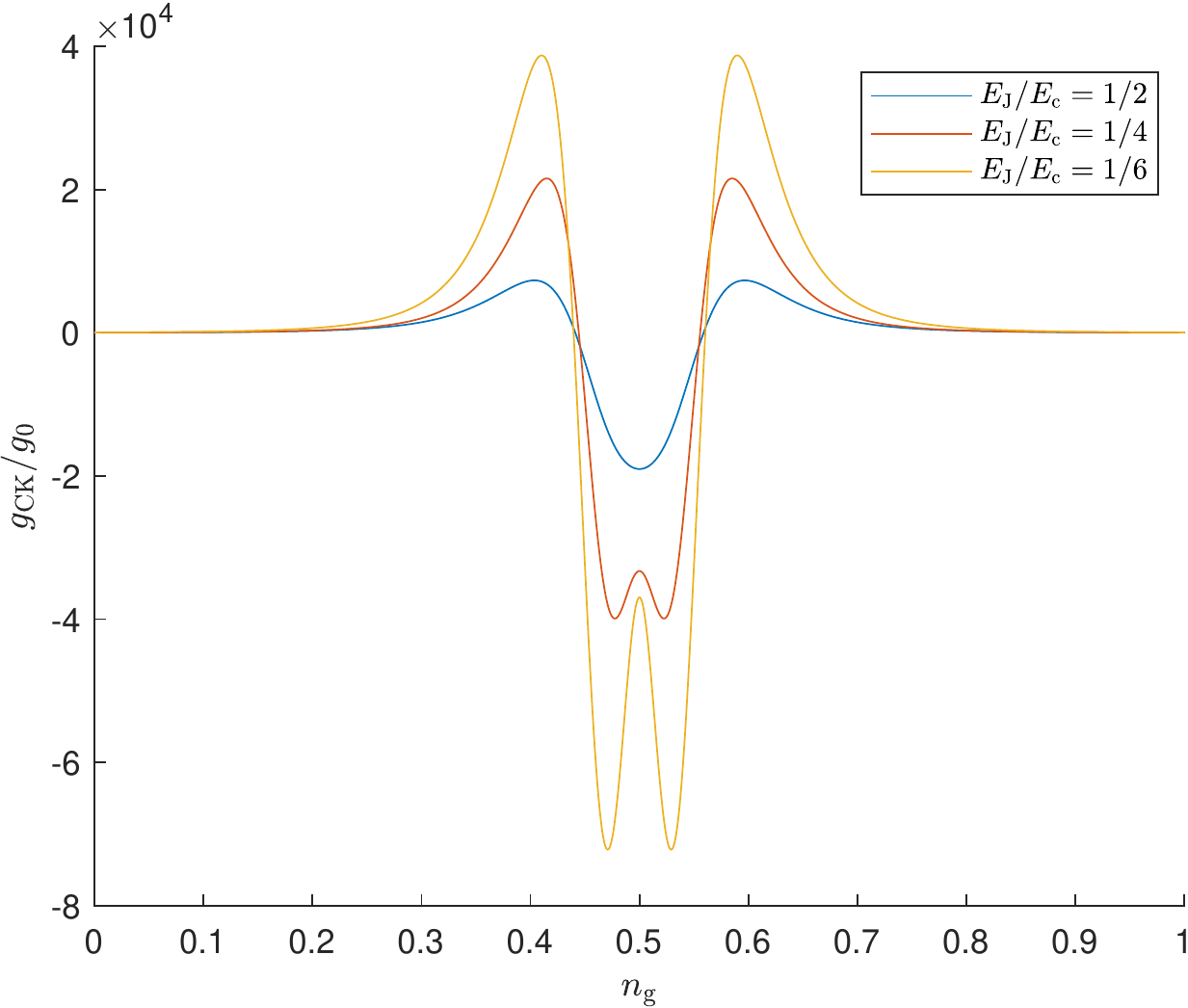}
        \caption{}
        \label{fig:gck_ng}
    \end{subfigure}
    \label{fig:gck_enhanced_full}
    \caption{\textbf{(a)} The cross-Kerr coupling $\gck$, scaled by $g_0$, arising from the circuit quantum capacitance calculations as a function of flux bias and gate charge. Here $\Ec/h = 30\,$GHz and $\Ej/h = 7.5\,$GHz. \textbf{(b)} The scaled cross-Kerr coupling for different $\Ej/\Ec$ ratios with $\Ej/h = 5\,$GHz at $\phie = 0$.}
\end{figure}

\twocolumngrid

\begin{widetext}
\begin{equation}
\label{eq:d2Ceffdx2}
\begin{split}
\frac{\partial^2 \Ceff}{\partial x^2} = & -\frac{\Cj \Cgone'^2}{\left( \Cgone + \Cj \right)^2} + \frac{\Cj \Cgone''}{\Cgone + \Cj} + \Cj \Cgone \left[ \frac{2 \Cgone'^2}{\left( \Cgone + \Cj \right)^3} - \frac{ \Cgone''}{\left( \Cgone + \Cj \right)^2} \right] \\
    &- \frac{1}{4 e^2} \left\{ \left( 2 \Cgone'^2 + 2 \Cgone \Cgone'' \right) \frac{\partial^2 E_k}{\partial \Ng^2}  - \left( 2 \Cgone \Cgone'^2 \frac{\Vg}{e} + \Cgone^2 \Cgone'' \frac{\Vg}{2 e} \right) \frac{\partial^3 E_k}{\partial \Ng^3} + \Cgone^2 \left( - \frac{\Vg}{2 e} \Cgone' \right)^2 \frac{\partial^4 E_k}{\partial \Ng^4} \right\} .
\end{split}
\end{equation}
\end{widetext}

Noticeably the radiation pressure coupling is antisymmetrical with respect to the degeneracy point $\Ng = 1/2$ whereas the cross-Kerr coupling is completely symmetrical in this sense and, additionally, the radiation pressure vanishes when the gate charge is tuned to the degeneracy point of the qubit, as shown in Figs. \ref{fig:grp_field_plot} and \ref{fig:grp_ng}. However, the cross-Kerr term does not vanish at this point, see Figs. \ref{fig:gck_field_plot} and \ref{fig:gck_ng}. By choosing $\Ej/\Ec \ll 1$ and detuning very close to charge degeneracy point, we are able to achieve a very strong Kerr nonlinearity without the optomechanical coupling. Therefore, with a proper choice of the gate charge and detuning, the system can be described with a simple Hamiltonian
\begin{equation}
\label{eq:simple_Kerr_Hamiltonian}
    \hat{\Hamil}_\mathrm{CK} = \hbar \wc \hat{a}^\dagger \hat{a} + \hbar \wm \hat{b}^\dagger \hat{b} + \hbar \gck \hat{a}^\dagger \hat{a} \hat{b}^\dagger \hat{b} .
\end{equation}

Looking at Eq. \eqref{eq:simple_Kerr_Hamiltonian}, the cavity resonant frequency is starkly shifted due to the number of phonons in the mechanical part, since with proper parameter selection the maximum predicted $\gck$ can reach up to typical microwave cavity linewidth $\kappa \sim 2 \pi \times 1 - 10\,$MHz.\cite{bib:Pirkkalainen2015,bib:Rochelau2010} This allows the system to function as a very good phonon counter with the cavity as the readout.

%xxxxxxxxxxxxxxxxxxxxxxxxxxxxxxxxxxxxxxxxxxxxxx
% PERTURBATIVE QM APPROACH
%xxxxxxxxxxxxxxxxxxxxxxxxxxxxxxxxxxxxxxxxxxxxxx
\section{Perturbative quantum mechanics approach}

Here we consider the dynamics of the system with a fully quantum mechanical treatment. This approach better highlights the involvement of the qubit to the enhancement of the cavity-mechanics coupling. In the above circuit model, the quantum capacitance of the Josephson junctions is seen to affect the total capacitance felt by the cavity, whereas here the coupling is seen to arise from a direct perturbation of the cavity and the mechanics on the qubit.

Naturally, the cavity and the mechanics can be considered as harmonic oscillators with Hamiltonians
\begin{subequations}
\begin{align}
    \hat{\Hamil}_\mathrm{c} &= \hbar \wc \hat{a}^\dagger \hat{a} , \label{eq:cavity_Hamiltonian_q} \\
    \hat{\Hamil}_\mathrm{m} &= \hbar \wm \hat{b}^\dagger \hat{b} . \label{eq:mechanics_Hamiltonian_q}
\end{align}
\end{subequations}
Here $\hat{a}^{(\dagger)}$ and $\hat{b}^{(\dagger)}$ are the annihilation (creation) operators of the cavity photons and the phonons in the moving capacitor, respectively, and we have neglected some terms arising from the detailed derivation provided in Appendix \ref{sec:quantization}, since these terms do not contribute to the radiation pressure and cross-Kerr couplings.

The charging energy Hamiltonian of the qubit in the charge basis, expressed with the number of Cooper pairs, is
\begin{equation}
\label{eq:charging_Hamiltonian}
    \hat{\Hamil}_\mathrm{ch} = 4 \Ec \sum_n \left[ \left( \hat{n} - \Ng \left( x \right) \right)^2 + \frac{2e}{\Csigmaonec} \hat{n} \hat{Q}_\mathrm{c} \right] \ket{n}\bra{n} .
\end{equation}
Here the first term is standard charging energy of a CPB and the second term causing a slight displacement of the qubit degeneracy point arises from the direct coupling between the cavity and the CPB. The effect that the second term has on the couplings turns out to be negligible with reasonable experimental parameters, but we include the term for completeness. We denote $\Csigmaonec = - 2 \Cc \left( \Cgone + \Cj \right)/\Cj$.

Considering the two lowest Cooper pair charge states $\ket{\mathrm{int}\left(\Ngzero\right)}$ and $\ket{\mathrm{int}\left(\Ngzero \right)+1}$ as the ground state and excited states, $\ket{0}$ and $\ket{1}$, respectively, the charging energy Hamiltonian becomes in the two-level system approximation
\begin{equation}
\label{eq:charging_Hamiltonian_2}
    \hat{\Hamil}_\mathrm{ch} = - \frac{B_3}{2} \sigma_z - \gcp \pc \sigma_z - \gm \xm \sigma_z + \gcm \xm \pc \sigma_z ,
\end{equation}
where the following shorthand notation is used
\begin{subequations}
\begin{align}
    \gm &= 4 \Ec \frac{\partial \Ng}{\partial x} \xzp = 4 \Ec \left( - \frac{1}{2e} \Cgone' \Vg \right) \xzp , \\
    \gcp &= 2 \frac{e}{\Csigmaonec} \Qzp \Ngzero , \\
    \gcm &= 2 \frac{e}{\Csigmaonec} \frac{\partial \Ng}{\partial x} \Qzp \xzp , \\
    \xm &= \hat{b} + \hat{b}^\dagger , \\
    \pc &= -i \left( \hat{a} - \hat{a}^\dagger \right) .
\end{align}
\end{subequations}

Here, $\sigma_x$, $\sigma_y$, $\sigma_z$ are Pauli matrices acting on the space spanned by the states $\ket{0}$ and $\ket{1}$, and $\hat{a}^{(\dagger)}$. The first term gives the unperturbed qubit excitation energy, and the second arises from qubit-cavity interaction, and the rest from the charge fluctuations $\delta \Ng = \frac{\partial \Cgone}{\partial x} \hat{x}$.

The node fluxes on the cavity side of the Josephson junctions and on the qubit island are $-\phic/2$ and $\phi_1$, respectively. The tunneling energy Hamiltonian is thus
\begin{equation}
\label{eq:Josephson_Hamiltonian}
\begin{split}
      \Hjj =& - \Ejone \cos \left( 2\pi \frac{\phi_1 + \frac{1}{2} \phic + \frac{1}{2} \phie}{\Phi_0} \right) \\
      &- \Ejtwo \cos \left( 2\pi \frac{\phi_1 + \frac{1}{2} \phic - \frac{1}{2} \phie}{\Phi_0} \right) \\
    =& \, - \Ej \cos \left( \pi \frac{\phie}{\Phi_0} \right) \left[ \cos \left( 2 \pi \frac{\phi_1}{\Phi_0} \right) \cos \left( \pi \frac{\phic}{\Phi_0} \right) \right. \\
    & \left. - \sin \left( 2 \pi \frac{\phi_1}{\Phi_0} \right) \sin \left( \pi \frac{\phic}{\Phi_0} \right) \right] ,
\end{split}
\end{equation}
where symmetrical JJs are assumed. The following results are all written assuming $\Ejone = \Ejtwo = \Ej/2$ for simplicity, and the full formulas are presented in the Appendices.

In general, a node flux can be related to the phase of the node with $\varphi_i = 2 \pi \frac{\phi_i}{\Phi_0}$. One can show that the quantized phase of the cavity node can be tied to the annihilation (creation) operator of photons $\hat{a}^{\left(\dagger\right)}$ so that $\hat{\varphi}_\mathrm{c}/2 = \eta (\hat{a}+\hat{a}^\dagger)$ where $\eta = \sqrt{e^2 Z_0/(2 \hbar)}$ with $Z_0 = \sqrt{\Lc/\Cc}$. The conjugate variable of the cavity flux, the cavity charge is similarly defined $\hat{Q}_\mathrm{c} = -i \Qzp \left( \hat{a} - \hat{a}^\dagger \right)$, where the zero-point motion of charge is $\Qzp = \sqrt{\frac{\hbar}{2 Z_0}}$. See Appendix \ref{sec:quantization} for the derivations.

The cavity-qubit coupling parameter $\eta \ll 1$ for typical microwave cavities. Therefore, we can now expand the sine and cosine terms of $\hat{\varphi}_\mathrm{c}$ in Eq. \eqref{eq:Josephson_Hamiltonian} up to second order in $\eta$. Properties of the phase operators also allow us to identify the superconducting phase of the island with ladder operators in the effective qubit so that considering on the states $\ket{0}$ and $\ket{1}$, we obtain $\cos(\hat{\varphi}_1 ) \mapsto \sigma_x /2$ and $\sin(\hat{\varphi}_1 ) \mapsto -\sigma_y /2$, see Appendix \ref{sec:quantization} for derivation. The approximate quantized Hamiltonian for the Josephson junctions is thus
\begin{equation}
\label{eq:Josephson_Hamiltonian_2}
\begin{split}
       \hat{\Hamil}_\mathrm{JJ} = - \frac{B_1}{2} \sigma_x - \frac{B_2}{2} \sigma_y + g_1 \sigma_y \xc + g_2 \sigma_x \xc^2
\end{split}
\end{equation}
with $g_1 = - B_1 \eta/2$, $g_2 = B_1 \eta^2/4$, and $\xc = \hat{a} + \hat{a}^\dagger$.

Let us decompose the Hamiltonians \eqref{eq:charging_Hamiltonian_2} and \eqref{eq:Josephson_Hamiltonian_2} into parts with the different Pauli matrices, and write the full system Hamiltonian
\begin{equation}
\label{eq:full_Hamiltonian}
\begin{split}
      \hat{\Hamil} = \hat{\Hamil}_\mathrm{c} + \hat{\Hamil}_\mathrm{m} - \frac{1}{2} \sum_{k=1}^3 \tilde{B}_k \sigma_k 
\end{split}
\end{equation}
with the perturbed qubit terms
\begin{subequations}
\begin{align}
    \tilde{B}_1 &= B_1 - 2 g_2 \xc^2 , \\
    \tilde{B}_2 &= - 2 g_1 \xc , \\
    \tilde{B}_3 &= B_3 + 2 \gm \xm + 2 \gcp \pc - 2 \gcm \pc \xm.
\end{align}
\end{subequations}
The Hamiltonian \eqref{eq:full_Hamiltonian} can thus be interpreted as the cavity and the mechanics slightly perturbing the isolated qubit parameters $B_k$ ($B_2 = 0$ for symmetrical JJs). In other words, the qubit mediates the interaction between the cavity and the mechanics leading to a notable enhancement of the coupling.

Let us consider a perturbative approach to solving the cavity-mechanics couplings from Eq. \eqref{eq:full_Hamiltonian}, and note that the eigenenergies of a qubit with Hamiltonian $- \frac{1}{2} \sum_{k=1}^3 \tilde{B}_k \sigma_k$ are $\pm \frac{1}{2} \sqrt{\tilde{B}_1^2 + \tilde{B}_2^2 + \tilde{B}_3^2}$. We write the Hamiltonian \eqref{eq:full_Hamiltonian} in terms of its ground state
\begin{equation}
\label{eq:full_Hamiltonian_2}
\begin{split}
    \hat{\Hamil} =& \hat{\Hamil}_\mathrm{c} + \hat{\Hamil}_\mathrm{m} - \frac{1}{2} \sqrt{\tilde{B}_1^2 + \tilde{B}_2^2 + \tilde{B}_3^2} \\
    =& \hat{\Hamil}_\mathrm{c} + \hat{\Hamil}_\mathrm{m} \\
    &-\frac{1}{2} B \Big\{ 1 + \frac{1}{B^2} \left( \beta \xc^2 + \delta \xc^4 + \epsilon \xm + \lambda \xm^2 \right. \\
    & \left. + \xi_1 \pc + \xi_2 \pc^2 + \xi_3 \pc \xm + \xi_4 \pc^2 \xm + \xi_5 \pc \xm^2 \right) \Big\}^{\frac{1}{2}}
\end{split}
\end{equation}
with
\begin{subequations}
\begin{align}
    B &= \sqrt{B_1^2 + B_3^2} , \\
    \beta &= 4 \left( g_1^2 - B_1 g_2 \right) , \\
    \delta &= 4 g_2^2 , \\
    \epsilon &= 4 B_3 \gm , \\
    \lambda &= 4 \gm^2 , \\
    \xi_1 &= 4 B_3 \gcp , \\
    \xi_2 &= 4 \gcp^2 , \\
    \xi_3 &= 4 \left( 2 \gcp \gm - B_3 \gcm \right) , \\
    \xi_4 &= -8 \gcp \gcm , \\
    \xi_5 &= - 8 \gm \gcm .
\end{align}
\end{subequations}
We can expand the square root term as $\sqrt{1+x} \approx 1 + \frac{x}{2} - \frac{x^2}{8}$, and we note that the expansion is more accurate further away from the degeneracy point of the qubit, where $B$ is larger. This technique yields cavity-mechanics couplings that approach the ones obtained from the circuit model far away from the degeneracy point when $\Ej \ll \Ec$. We also note that the terms $\xi_i$ are negligible compared to the other contributions using reasonable experimental parameters, and expanding the Hamiltonian to higher orders is easier when these terms are omitted after which all terms in the expanded Hamiltonian commute.

Up to the second order in the expansion, we obtain
\begin{equation}
\label{eq:sqrt_expansion}
\begin{split}
    & - \frac{1}{2} \sqrt{\tilde{B}_1^2 + \tilde{B}_2^2 + \tilde{B}_3^2} \\
    \approx & - \frac{1}{2} B \\
    & - \frac{1}{4 B} \left( \beta \xc^2 + \delta \xc^4 + \epsilon \xm + \lambda \xm^2 + \xi_1 \pc \right. \\
    & \ \ \ \ \ \ \ \ \ \left.  + \xi_2 \pc^2 + \xi_3 \pc \xm + \xi_4 \pc^2 \xm + \xi_5 \pc \xm^2 \right) \\
    & + \frac{1}{16 B^3} \left( \beta \xc^2 + \delta \xc^4 + \epsilon \xm + \lambda \xm^2  + \xi_1 \pc \right. \\
    & \ \ \ \ \ \ \ \ \ \left.  + \xi_2 \pc^2 + \xi_3 \pc \xm + \xi_4 \pc^2 \xm + \xi_5 \pc \xm^2 \right)^2,
\end{split}
\end{equation}
where, after normal ordering the cavity and mechanics operators, we identify the radiation pressure coupling as the prefactor of the term $-\hbar \grp \hat{a}^\dagger \hat{a} \left( \hat{b}^\dagger + \hat{b} \right)$
\begin{equation}
\label{eq:SWT_grp}
\begin{split}
    \hbar \grp = & \frac{1}{2 B} \xi_4 - \frac{1}{4 B^3} \left[ \epsilon \left( \beta + 6 \delta + \xi_2 \right) \right. \\
    &\left. + \xi_4 \left( 2 \beta - 3 \delta + 3 \lambda \right) \right]
\end{split}
\end{equation}
and the cross-Kerr coupling as the prefactor of the term $\hbar \gck \hat{a}^\dagger \hat{a} \hat{b}^\dagger \hat{b}$
\begin{equation}
\label{eq:SWT_gck}
\begin{split}
    \hbar \gck =& \frac{1}{4 B^3} \left[ 2 \lambda \left( \beta + 6 \delta \right) + \xi_3^2 + 6 \xi_4^2 + 6 \xi_5^2 \right. \\
    & \left.+ 2 \epsilon \xi_4 + 2 \lambda \xi_4 \right] .
\end{split}
\end{equation}
The circuit and quantum mechanical descriptions of the system dynamics agree qualitatively but have differences quantitatively close to the degeneracy point of the qubit due to the inaccuracy of the perturbative quantum mechanical approach in that regime. However, importantly the circuit and quantum mechanical models align well far away from the degeneracy point and better agreement is naturally obtained by going to higher orders in the expansion of the $- \frac{1}{2} \sqrt{\tilde{B}_1^2 + \tilde{B}_2^2 + \tilde{B}_3^2}$ term. In Appendix \ref{sec:appendix_QM_perturbation}, we calculate the third order results for these couplings by omitting the $\xi_i$ terms that are small compared to other terms in the expansion. An important agreement of the two descriptions is also that at the degeneracy point the radiation pressure coupling vanishes and the cross-Kerr coupling obtains a large nonzero value enabling a rich platform for phonon counting experiments.

%xxxxxxxxxxxxxxxxxxxxxxxxxxxxxxxxxxxxxxxxxxxxxx
% DISCUSSION
%xxxxxxxxxxxxxxxxxxxxxxxxxxxxxxxxxxxxxxxxxxxxxx
\section{Discussion}

We are interested in reaching the ultrastrong single-photon coupling regime $\grp > \wm$ where one can observe the intrinsic nonlinearity of optomechanical coupling that goes unseen for weaker couplings that allow the linearization of the interaction. With our scheme, we are able to obtain radiation pressure coupling enhancement of seven orders of magnitude by utilizing the high range of tunability offered by the setup. With the proper selection of gate charge and the magnetic flux through the qubit loop, the desired coupling strength for a specific purpose can be obtained. Moreover, owing to the wide tuning options, one can find a regime with enhanced radiation pressure coupling and vanishing cross-Kerr coupling or vice versa which makes this setup practical for multiple types of studies.

For an optomechanical setup with a direct radiation pressure coupling $g_0$ of the order of $10\,$Hz, we are thus able to reach a coupling of the order of $100\,$MHz, which facilitates probing of the ground state of a typical flexural nanomechanical resonator without the need for sideband cooling.

Above, we performed our model calculations for a  cavity with impedance $Z_0 = 100\,\Omega$, which means  that the zero-point fluctuations of flux $\phizp = \sqrt{\hbar Z_0/2}$ exceed the zero-point fluctuation of charge $\Qzp = \sqrt{\hbar/\left(2 Z_0 \right)}$, making the device characteristics more strongly influenced by flux than charge fluctuations. As seen in Figs. \ref{fig:grp_field_plot} and \ref{fig:gck_field_plot}, the enhancement of the radiation pressure and cross-Kerr couplings is more affected by changes in $\Ng$ than in $\phie$. With proper tuning of $\Ng$, sizeable enhancements to $\grp$ and $\gck$ can still be obtained even with averaging effects in $\phie$ arising from large phase fluctuations.

In summary, we have analyzed an optomechanical setup based on utilization of Josephson (“quantum”) capacitance of a Cooper-pair box to boost the optomechanical coupling. We reach an enhancement of radiation pressure by seven orders of magnitude, which brings the system well into the ultrastrong coupling regime. The coupling is highly tunable by charge and flux bias and, by proper selection of the bias point, strongly enhanced Cross-Kerr coupling without radiation pressure effects can be achieved for phonon counting purposes.

\section*{Acknowledgments}

We thank Andreas H\"uttel, Yuriy Makhlin, Francesco Massel, and Tero Heikkil\"a for useful discussions. This work was supported by the Academy of Finland Center of Excellence ``Quantum Technology Finland'' (Grant No. 312295) and RADDESS program (Grant No. 314448, BOLOSE), as well as by the European Research Council (Grant No. 670743). Mohammad Tasnimul Haque acknowledges support from the European Union’s Horizon 2020 Programme for Research and Innovation under grant agreement No. 722923 (Marie Curie ETN - OMT) and a three-month-long secondment at University of Camerino, during which  part of the work was carried out. Juuso Manninen thanks the support of the Vilho, Yrj{\"o} and Kalle V{\"a}is{\"a}l{\"a} Foundation of the Finnish Academy of Science and Letters. This work was also supported in part by COST Action CA16218 (NANOCOHYBRI) and the European Microkelvin Platform (EMP, No. 824109).

%XXXXXXXXXXXXXXXXXXXXXXXXXXXXXXXXXXXXXX
% APPENDICES
%XXXXXXXXXXXXXXXXXXXXXXXXXXXXXXXXXXXXXX

\onecolumngrid
%\newpage
\appendix
%\section{Details of Some Calculations}
%includepdf[pages={1-4}]{0958_001.pdf}
%\nocite{*}

%XXXXXXXXXXXXXXXXXXXXXXXXXXXXXXXXXXXXXX
% Forming the Hamiltonian
%XXXXXXXXXXXXXXXXXXXXXXXXXXXXXXXXXXXXXX
\section{Forming the Hamiltonian}

\begin{figure}[htb]
  \includegraphics[width=0.9\linewidth]{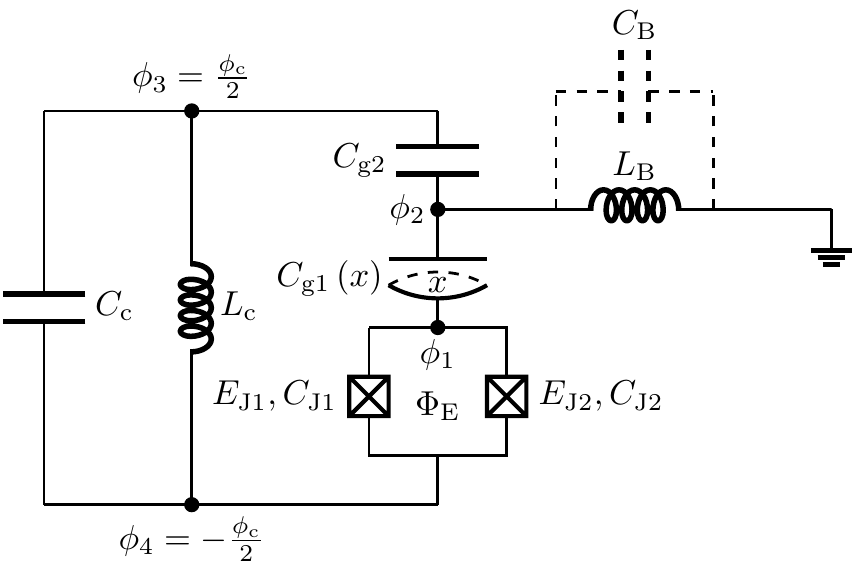}
  \caption{Schematics of the Cooper-pair box circuit for enhancement of optomechanical coupling. For elements, see text.}
  \label{fig:CSET_schematic}
\end{figure}

We define the node flux at the node $i$ at time $t$ as
\begin{equation}
\label{eq:phi_definition}
    \phi_i \left( t \right) = \int^t V_i \left( \tau \right) \, \mathrm{d}\tau .
\end{equation}
This implies that the voltage at node $i$ can be expressed as
\begin{equation}
\label{eq:voltage_definition}
    V_i \left( t \right) = \dot{\phi}_i .
\end{equation}
The node flux is also related to the phase of the node with
\begin{equation}
\label{eq:flux_phase_relation}
    \varphi_i = 2 \pi \frac{\phi_i}{\Phi_0} ,
    %\varphi_i = \frac{\phi_i}{\Phi_0} ,
\end{equation}
where %$\Phi_0 = \frac{\hbar}{2e}$
$\Phi_0 = \frac{h}{2e}$ is the flux quantum. The node indices corresponding to each capacitive island of the circuit can be seen in Fig. \ref{fig:CSET_schematic}.

The energy stored in the capacitive elements of the circuit is
\begin{equation}
\label{eq:capacitive_energy}
\begin{split}
    \kinen &= \frac{\Cj}{2} \left( \phionedot - \phifourdot \right)^2 + \frac{\Cgone}{2} \left( \phionedot - \phitwodot \right)^2 + \frac{\Cgtwo}{2} \left( \phitwodot - \phithreedot \right)^2 + \frac{\Cc}{2} \left( \phithreedot^2 - \phifourdot^2 \right)^2 + \frac{\Cb}{2} \phitwodot^2 \\
    &= \frac{\Cj}{2} \left( \phionedot + \frac{1}{2} \phicdot \right)^2 + \frac{\Cgone}{2} \left( \phionedot - \phitwodot \right)^2 + \frac{\Cgtwo}{2} \left( \phitwodot - \frac{1}{2} \phicdot \right)^2 + \frac{\Cc}{2} \phicdot^2 + \frac{\Cb}{2} \phitwodot^2 ,
\end{split}
\end{equation}
where $\Cj = \Cjone + \Cjtwo$ is the capacitance of the Josephson junctions, $\Cgone$ and $\Cgtwo$ the gate capacitances, and $\Cc$ the capacitance of the LC cavity. The position dependence of $\Cgone = \Cgone \left( x \right)$ is omitted for notational convenience. The bias capacitance $\Cb$ is included to induce a bias voltage on to the island 2. After taking $\Cb \rightarrow \infty$, the voltage on island 2 corresponds to the bias voltage. Detailed calculations are given below. Additionally, in order for the method of nodes to work, the net of the capacitive elements connecting all of the nodes of the circuit needs to be simply connected. \cite{bib:Vool2017}

Similarly, the inductive energy of the system is
\begin{equation}
\label{eq:inductive_energy}
\begin{split}
    \poten =&\, - \Ejone \cos \left( 2\pi \frac{\phi_1 - \phi_4 + \frac{1}{2} \phie}{\Phi_0} \right) - \Ejtwo \cos \left( 2\pi \frac{\phi_1 - \phi_4 - \frac{1}{2} \phie}{\Phi_0} \right) \\
    &+ \frac{1}{2 \Lb} \phi_2^2 + \frac{1}{2 \Lc} \left( \phi_3 - \phi_4 \right)^2 \\
    =&\, - \Ejone \cos \left( 2\pi \frac{\phi_1 + \frac{1}{2} \phic + \frac{1}{2} \phie}{\Phi_0} \right) - \Ejtwo \cos \left( 2\pi \frac{\phi_1 + \frac{1}{2} \phic - \frac{1}{2} \phie}{\Phi_0} \right) \\
    &+ \frac{1}{2 \Lb} \phi_2^2 + \frac{1}{2 \Lc} \phic^2 .
\end{split}
\end{equation}
Here $\Ejone$, $\Ejtwo$ are the Josephson energies of the junctions, $\phie$ the external flux through the loop of the Cooper-pair box, $\Lb$ the bias inductance, and $\Lc$ the inductance of the LC cavity.

The Lagrangian $\Lagr \left( \phi_i, \dot{\phi}_i \right)$ is thus
\begin{equation}
\label{eq:Lagrangian}
    \Lagr = \kinen - \poten = \frac{1}{2} \vec{\dot{\phi}}^\intercal \left[ C \right] \vec{\dot{\phi}} - \poten,
\end{equation}
where
\begin{equation}
\label{eq:capacitance_matrix}
    \left[ C \right] =\begin{pmatrix}
                        \Cj + \Cgone & -\Cgone & \frac{1}{2} \Cj \\
                        -\Cgone & \Cgone + \Cgtwo + \Cb & -\frac{1}{2} \Cgtwo \\
                        \frac{1}{2} \Cj & - \frac{1}{2} \Cgtwo & \Cc + \frac{1}{4} \Cj + \frac{1}{4} \Cgtwo 
                        \end{pmatrix}
\hspace{0.5cm} ; \hspace{0.5cm}
    \vec{\dot{\phi}} = \begin{pmatrix}
                        \phionedot \\
                        \phitwodot \\
                        \phicdot
                        \end{pmatrix} .
\end{equation}

Now, the Hamiltonian of the system $\Hamil \left( \phi_i , Q_i \right)$ can be expressed with the conjugate momenta %
\begin{equation}
\label{eq:conjugate_momenta}
    Q_i = \frac{\partial \Lagr}{\partial \dot{\phi}_i}
\end{equation}
that correspond to the electric charge on the island $i$. The canonical relation between the Lagrangian and the Hamiltonian gives
\begin{equation}
\label{eq:Hamiltonian}
\begin{split}
    \Hamil &= \sum_i \dot{\phi}_i \frac{\partial \Lagr}{\partial \dot{\phi}_i} - \Lagr = \sum_i \dot{\phi}_i Q_i - \Lagr \\
    &= \frac{1}{2} \vec{Q}^\intercal \left[ C \right]^{-1} \vec{Q} + \poten .
\end{split}
\end{equation}
Here, the inverse of the capacitance matrix is
\begin{equation}
\label{eq:inverse_capacitance_matrix}
    \left[ C \right]^{-1} = \begin{pmatrix}
                            \frac{1}{\Csigmaone} & \frac{1}{\Csigmaonetwo} &  \frac{1}{\Csigmaonec} \\
                            \frac{1}{\Csigmaonetwo} & \frac{1}{\Csigmatwo} &  \frac{1}{\Csigmatwoc} \\
                            \frac{1}{\Csigmaonec} & \frac{1}{\Csigmatwoc} & \frac{1}{\Csigmac}
                            \end{pmatrix}
\end{equation}
with the following shorthand notations
\begin{subequations}
\label{eq:shorthand_notations}
\begin{align}
    \frac{1}{\Csigmaone} &= \frac{\Cgone \Cgtwo + 4 \Cc \left( \Cgone + \Cgtwo\right) + \Cj \left( \Cgone + \Cgone \right) + \Cb \left( 4 \Cc + \Cgtwo + \Cj \right)}{\Cb \Cgone \left( 4 \Cc + \Cgtwo \right) + \Cb \Cj \left( 4 \Cc + \Cgone + \Cgtwo \right) + 4 \left[ \Cc \Cgone \Cgtwo + \Cj \Cgone \Cgtwo + \Cc \Cj \left( \Cgone + \Cgtwo \right) \right]}, \\
    \frac{1}{\Csigmatwo} &= \frac{\Cgtwo \Cj + 4 \Cc \left( \Cgone + \Cgtwo \right) + \Cgone \left( \Cgtwo + \Cj \right)}{\Cb \Cgone \left( 4 \Cc + \Cgtwo \right) + \Cb \Cj \left( 4 \Cc + \Cgone + \Cgtwo \right) + 4 \left[ \Cc \Cgone \Cgtwo + \Cj \Cgone \Cgtwo + \Cc \Cj \left( \Cgone + \Cgtwo \right) \right]}, \\
    \frac{1}{\Csigmac} &= \frac{4 \Cgone \left( \Cb + \Cgtwo \right) + 4 \Cj \left( \Cb + \Cgone + \Cgtwo \right)}{\Cb \Cgone \left( 4 \Cc + \Cgtwo \right) + \Cb \Cj \left( 4 \Cc + \Cgone + \Cgtwo \right) + 4 \left[ \Cc \Cgone \Cgtwo + \Cj \Cgone \Cgtwo + \Cc \Cj \left( \Cgone + \Cgtwo \right) \right]}, \\
    \frac{1}{\Csigmaonetwo} &= \frac{4 \Cc \Cgone - \Cgtwo \Cj + \Cgone \left( \Cgtwo + \Cg \right)}{\Cb \Cgone \left( 4 \Cc + \Cgtwo \right) + \Cb \Cj \left( 4 \Cc + \Cgone + \Cgtwo \right) + 4 \left[ \Cc \Cgone \Cgtwo + \Cj \Cgone \Cgtwo + \Cc \Cj \left( \Cgone + \Cgtwo \right) \right]}, \\
    \frac{1}{\Csigmaonec} &= \frac{2 \Cgone \Cgtwo - 2 \Cj \left( \Cb + \Cgone + \Cgtwo \right)}{\Cb \Cgone \left( 4 \Cc + \Cgtwo \right) + \Cb \Cj \left( 4 \Cc + \Cgone + \Cgtwo \right) + 4 \left[ \Cc \Cgone \Cgtwo + \Cj \Cgone \Cgtwo + \Cc \Cj \left( \Cgone + \Cgtwo \right) \right]}, \\
    \frac{1}{\Csigmatwoc} &= \frac{2 \Cgone \left( \Cgtwo - \Cj \right) + \Cgtwo \Cj}{\Cb \Cgone \left( 4 \Cc + \Cgtwo \right) + \Cb \Cj \left( 4 \Cc + \Cgone + \Cgtwo \right) + 4 \left[ \Cc \Cgone \Cgtwo + \Cj \Cgone \Cgtwo + \Cc \Cj \left( \Cgone + \Cgtwo \right) \right]} .
\end{align}
\end{subequations}
Thus, the Hamiltonian can be written out explicitly
\begin{equation}
\label{eq:Hamiltonian_explicit}
    \Hamil = \frac{1}{2 \Csigmaone} Q_1^2 + \frac{1}{2 \Csigmatwo} Q_2^2 + \frac{1}{2 \Csigmac} Q_\mathrm{c}^2 + \frac{1}{\Csigmaonetwo} Q_1 Q_2 + \frac{1}{\Csigmaonec} Q_1 Q_\mathrm{c} + \frac{1}{\Csigmatwoc} Q_2 Q_\mathrm{c} + \poten .
\end{equation}
Define the nominal bias voltage
\begin{equation}
\label{eq:Vg_definition}
    \Vg = \frac{Q_2}{\Csigmatwo}
\end{equation}
and calculate the real voltage on island 2
\begin{equation}
\label{eq:island_2_voltage}
    \frac{\partial \Hamil}{\partial Q_2} = \frac{1}{\Csigmatwo} Q_2 + \frac{1}{\Csigmaonetwo} Q_1 + \frac{1}{\Csigmatwoc} Q_\mathrm{c} = \Vg + \frac{1}{\Csigmaonetwo} Q_1 + \frac{1}{\Csigmatwoc} Q_\mathrm{c} .
\end{equation}
In the limit $\Cb \rightarrow \infty$, we obtain $\frac{\partial \Hamil}{\partial Q_2} = \Vg$, i.e. island 2 is now set to a potential that can be tuned with $\Vg$.

Define the charging energy of the Cooper-pair box, the number of Cooper pairs on island 1, and the nominal gate charge
\begin{subequations}
\label{eq:CPB_equations}
\begin{align}
    \Ec &= \frac{e^2}{2 \Csigmaone}, \label{eq:charging_energy} \\
    n &= \frac{Q_1}{2 e}, \label{eq:number_of_Cooper_pairs} \\
    \Ng &= - \frac{1}{2e} \frac{\Csigmaone \Csigmatwo}{\Csigmaonetwo} \Vg .
\end{align}
\end{subequations}
With these definitions along with Eq. \eqref{eq:Vg_definition}, the Hamiltonian \eqref{eq:Hamiltonian_explicit} can be reformulated to a standard form
\begin{equation}
\label{eq:Hamiltonian_standard_form}
\begin{split}
    \Hamil &= 4 \Ec \left( n - \Ng \right)^2 + 4 \Ec \left( \frac{\Csigmaonetwo^2}{\Csigmaone \Csigmatwo} - 1 \right) \Ng^2 \\
    &+ \frac{1}{2 \Csigmac} Q_\mathrm{c}^2 + \left( \frac{1}{\Csigmaonec} \cdot 2 e n + \frac{\Csigmaonetwo}{\Csigmaone \Csigmatwoc} \cdot 2 e \Ng \right) Q_\mathrm{c} + \poten .
\end{split}
\end{equation}
Let us now focus on $\poten$ in Eq. \eqref{eq:inductive_energy} and specifically the Josephson junction part of it. Let us define the Josephson energies of the junctions as
\begin{subequations}
\label{eq:Josephson_energies}
\begin{align}
    \Ejone &= \left( 1+d \right) \frac{\Ej}{2} , \\
    \Ejtwo &= \left( 1-d \right) \frac{\Ej}{2} ,
\end{align}
\end{subequations}
where $d \in \left[ \left. 0,1 \right. \right)$ describes the difference of the Josephson energies of the two junctions. Using Eq. \eqref{eq:Josephson_energies} and the identity $\cos \left( \alpha \pm \beta \right) = \cos \alpha \cos \beta \mp \sin \alpha \sin \beta$, we can rewrite
\begin{equation}
\label{eq:Josephson_inductances}
\begin{split}
    &- \Ejone \cos \left( 2\pi \frac{\phi_1 + \frac{1}{2} \phic + \frac{1}{2} \phie}{\Phi_0} \right) - \Ejtwo \cos \left( 2\pi \frac{\phi_1 + \frac{1}{2} \phic - \frac{1}{2} \phie}{\Phi_0} \right) \\
    =& \, - \Ej \cos \left( \pi \frac{\phie}{\Phi_0} \right) \left[ \cos \left( 2 \pi \frac{\phi_1}{\Phi_0} \right) \cos \left( \pi \frac{\phic}{\Phi_0} \right) - \sin \left( 2 \pi \frac{\phi_1}{\Phi_0} \right) \sin \left( \pi \frac{\phic}{\Phi_0} \right) \right] \\
    &+ \Ej d \sin \left( \pi \frac{\phie}{\Phi_0} \right) \left[ \cos \left( 2 \pi \frac{\phi_1}{\Phi_0} \right) \sin \left( \pi \frac{\phic}{\Phi_0} \right) + \sin \left( 2 \pi \frac{\phi_1}{\Phi_0} \right) \cos \left( \pi \frac{\phic}{\Phi_0} \right) \right] .
\end{split}
\end{equation}
Now, the total Hamiltonian of the system can be written as
\begin{equation}
\label{eq:total_Hamiltonian}
\begin{split}
    \Hamil =& 4 \Ec \left( n - \Ng \right)^2 + 4 \Ec \left( \frac{\Csigmaonetwo^2}{\Csigmaone \Csigmatwo} - 1 \right) \Ng^2 \\
    &+ \frac{1}{2 \Csigmac} Q_\mathrm{c}^2 + \left( \frac{1}{\Csigmaonec} \cdot 2 e n + \frac{\Csigmaonetwo}{\Csigmaone \Csigmatwoc} \cdot 2 e \Ng \right) Q_\mathrm{c} \\
    &- \Ej \cos \left( \pi \frac{\phie}{\Phi_0} \right) \left[ \cos \left( 2 \pi \frac{\phi_1}{\Phi_0} \right) \cos \left( \pi \frac{\phic}{\Phi_0} \right) - \sin \left( 2 \pi \frac{\phi_1}{\Phi_0} \right) \sin \left( \pi \frac{\phic}{\Phi_0} \right) \right] \\
    &+ \Ej d \sin \left( \pi \frac{\phie}{\Phi_0} \right) \left[ \cos \left( 2 \pi \frac{\phi_1}{\Phi_0} \right) \sin \left( \pi \frac{\phic}{\Phi_0} \right) + \sin \left( 2 \pi \frac{\phi_1}{\Phi_0} \right) \cos \left( \pi \frac{\phic}{\Phi_0} \right) \right] \\
    &+ \frac{1}{2 \Lb} \phi_2^2 + \frac{1}{2 \Lc} \phic^2 .
\end{split}
\end{equation}
In the limit $\Cb \rightarrow \infty$, the definitions introduced in Eqs. \eqref{eq:shorthand_notations} and \eqref{eq:CPB_equations} simplify to
\begin{subequations}
\label{eq:definitions_infinite_Cb_limit}
\begin{align}
    \frac{1}{\Csigmaone} &\rightarrow \frac{4 \Cc + \Cgtwo + \Cj}{\Cgone \left( 4 \Cc + \Cgtwo \right) + \Cj \left( 4 \Cc + \Cgone + \Cgtwo \right)} \approx \frac{1}{\Cgone + \Cj} , \\
    \frac{1}{\Csigmatwo} &\rightarrow 0 , \\
    \frac{1}{\Csigmac} &\rightarrow \frac{4 \Cgone + 4 \Cj}{\Cgone \left( 4 \Cc + \Cgtwo \right) + \Cj \left( 4 \Cc + \Cgone + \Cgtwo \right)} \approx \frac{1}{\Cc} , \\
    \frac{1}{\Csigmaonec} &\rightarrow - \frac{2 \Cj}{\Cgone \left( 4 \Cc + \Cgtwo \right) + \Cj \left( 4 \Cc + \Cgone + \Cgtwo \right)} \approx - \frac{\Cj}{2 \Cc \left( \Cgone + \Cj \right)} , \\
    \frac{1}{\Csigmaonetwo} &\rightarrow 0 , \\
    \frac{1}{\Csigmatwoc} &\rightarrow 0 , \\
    \Ng &\rightarrow - \left( \Cgone - \frac{\Cgtwo \Cj}{4 \Cc + \Cgtwo + \Cj} \right) \frac{\Vg}{2e} \approx - \frac{\Cgone}{2 e} \Vg . \label{eq:gate_charge_infinite_Cb_limit}
\end{align}
\end{subequations}
The rightmost approximate results are recovered in the case, where $\Cc$ is the dominant capacitance in the system.

%XXXXXXXXXXXXXXXXXXXXXXXXXXXXXXXXXXXXXX
% Deriving the effective capacitance
%XXXXXXXXXXXXXXXXXXXXXXXXXXXXXXXXXXXXXX
\section{Deriving the effective capacitance}

In the following we determine the effective capacitance of the Cooper-pair box part of the circuit. Let us first determine the electric charges on the different islands of the circuit. These arise from the relation in Eq. \eqref{eq:conjugate_momenta}
\begin{subequations}
\label{eq:island_charges}
\begin{align}
    Q_1 &= \frac{\partial \Lagr}{\partial \phionedot} = \left( \Cj + \Cgone \right) \phionedot - \Cgone \phitwodot + \frac{1}{2} \Cj \phicdot , \label{eq:Q1}\\
    Q_2 &= \frac{\partial \Lagr}{\partial \phitwodot} = \left( \Cgone + \Cgtwo + \Cb \right) \phitwodot - \Cgone \phionedot - \frac{1}{2} \Cgtwo \phicdot , \label{eq:Q2}\\
    Q_\mathrm{c} &= \frac{\partial \Lagr}{\partial \phithreedot} = \left( \Cc + \frac{1}{4} \Cgtwo + \frac{1}{4} \Cj \right) \phicdot + \frac{1}{2} \Cj \phitwodot - \frac{1}{2} \Cgtwo \phitwodot \label{eq:Q3}.
\end{align}
\end{subequations}
Recall the definition of $\Vg$ in Eq. \eqref{eq:Vg_definition}. In the large $\Cb$ limit, $\Csigmatwo \rightarrow \Cb$, and together with the relation \eqref{eq:Q2}, $\phitwodot = \Vg$ is implied. Now that the voltage on the island is fixed to $\Vg$ in this limit, the number of Cooper pairs on island 1 can be determined by combining the relations \eqref{eq:number_of_Cooper_pairs} and \eqref{eq:Q1}
\begin{equation}
\label{eq:number_of_Cooper_pairs_2}
    n = \frac{Q_1}{2 e} = \frac{ \left( \Cj + \Cgone \right) \phionedot - \Cgone \Vg + \frac{1}{2} \Cj \phicdot}{2 e}.
\end{equation}
The voltage on the Cooper-pair box island is thus
\begin{equation}
\label{eq:CPB_voltage}
    \Vi = \phionedot = \frac{\Cgone}{\Cj + \Cgone} \Vg - \frac{1}{2} \frac{\Cj}{\Cj + \Cgone} \phicdot + \frac{2 e n}{\Cj + \Cgone}
\end{equation}
and, therefore, the charge across $\Cgone$ can be written
\begin{equation}
\label{eq:charge_across_gate}
    Q_{\mathrm{g}1} = \Cgone \left( \Vg - \Vi \right) = \frac{\Cj \Cgone}{\Cj + \Cgone} \Vg - \frac{\Cgone}{\Cj + \Cgone} 2 e n + \frac{1}{2} \frac{\Cj \Cgone}{\Cj + \Cgone} \phicdot .
\end{equation}
Notice that here the second term has the opposite sign compared to the calculation presented by Duty et al. \cite{bib:Duty2005}. However, also our gate charge is defined with the opposite sign with respect to the gate voltage, see Eq. \eqref{eq:gate_charge_infinite_Cb_limit}. Thus the effective capacitance obtained here aligns with the results in \cite{bib:Duty2005}
\begin{equation}
\label{eq:effective_capacitance_appendix}
    \Ceff = \frac{\partial Q_{\mathrm{g}1}}{\partial \Vg} = \frac{\Cgone \Cj}{\Csigmaone} - \frac{\Cgone^2}{4 e^2} \frac{\partial^2 E_k}{\partial \Ng^2} ,
\end{equation}
where $E_k$ is the $k$th energy band of the Cooper-pair box. We do not have to take the position dependence of $\Cgone$ into consideration in this part of the derivation. Since were are considering a voltage bias setup with a resonance frequency well below $RC$ cutoff effects, the voltage on the Cooper-pair box island is able to follow the movement of the capacitor without difficulties.

%XXXXXXXXXXXXXXXXXXXXXXXXXXXXXXXXXXXXXX
% Radiation pressure and cross-Kerr couplings
%XXXXXXXXXXXXXXXXXXXXXXXXXXXXXXXXXXXXXX
\section{Radiation pressure and cross-Kerr couplings}

Consider the resonance frequency of an LC circuit
\begin{equation}
\label{eq:resonance_frequency}
    \wc = \frac{1}{\sqrt{\Lc \Ctot}} , 
\end{equation}
where the total capacitance $\Ctot$ arises from the cavity capacitor that is parallel with the second gate capacitor and the effective Cooper-pair box capacitor
\begin{equation}
\label{eq:total_capacitance}
    \Ctot = \Cc + \left( \frac{1}{\Cgtwo} + \frac{1}{\Ceff} \right)^{-1} .
\end{equation}
The cavity frequency can be expanded in the position of the moving capacitor $\Cgone \left( x \right)$
\begin{equation}
\label{eq:resonance_frequency_expansion}
    \wc \simeq \omega_{\mathrm{c}0} + \frac{\partial \wc}{\partial x} x + \frac{1}{2} \frac{\partial^2 \wc}{\partial x^2} x^2 .
\end{equation}
Here, the linear term corresponds to the radiation pressure coupling. Thinking about a simple optical cavity, the decrease in cavity length should lead to the increase in cavity frequency. The quantization of the circuit gives the position operator the form $\hat{x} = \xzp \left( b^\dagger + b \right)$, where $\xzp$ is the zero point motion. Thus the radiation pressure coupling is
\begin{equation}
\label{eq:grp}
    \hbar \grp = - \hbar \frac{\partial \wc}{\partial x} \xzp .
\end{equation}
A straightforward calculation yields
\begin{equation}
\label{eq:dwcdx}
    \frac{\partial \wc}{\partial x} = - \frac{1}{2} \frac{\Cgtwo^2}{\left( \Cgtwo + \Ceff \right)^2} \frac{\wc}{\Ctot} \frac{\partial \Ceff}{\partial x}
\end{equation}
with
\begin{equation}
\label{eq:dCeffdx}
    \frac{\partial \Ceff}{\partial x} = -\frac{\Cj \Cgone \Cgone'}{\left(\Cgone + \Cj\right)^2} + \frac{\Cj \Cgone'}{\Cgone + \Cj} - \frac{\Cgone \Cgone'}{2 e^2} \frac{\partial^2 E_k}{\partial \Ng^2} - \frac{\Cgone^2}{4 e^2} \frac{\partial}{\partial x} \left( \frac{\partial^2 E_k}{\partial \Ng^2} \right)
\end{equation}
arising from the expression \eqref{eq:effective_capacitance}. Here the prime notation is used to mark the x-derivatives. Using the chain rule, we can determine
\begin{equation}
\label{eq:dx_energy_band}
    \frac{\partial}{\partial x} \left( \frac{\partial^2 E_k}{\partial \Ng^2} \right) = \frac{\partial \Ng}{\partial x} \frac{\partial}{\partial \Ng} \left( \frac{\partial^2 E_k}{\partial \Ng^2} \right) = -\frac{\Vg}{2 e} \Cgone' \frac{\partial^3 E_k}{\partial \Ng^3} ,
\end{equation}
where the gate charge definition \eqref{eq:gate_charge_infinite_Cb_limit} is used. Thus we obtain 
\begin{equation}
\label{eq:dCeffdx_2}
    \frac{\partial \Ceff}{\partial x} = -\frac{\Cj \Cgone \Cgone'}{\left(\Cgone + \Cj\right)^2} + \frac{\Cj \Cgone'}{\Cgone + \Cj} - \frac{\Cgone \Cgone'}{2 e^2} \frac{\partial^2 E_k}{\partial \Ng^2} + \frac{\Cgone^2 \Cgone'}{8 e^3} \Vg \frac{\partial^3 E_k}{\partial \Ng^3}
\end{equation}
leading to the radiation pressure coupling
\begin{equation}
\label{eq:grp_2}
\begin{split}
        \hbar \grp =& \hbar \xzp \left[ \frac{1}{2} \frac{\Cgtwo^2}{\left( \Cgtwo + \Ceff \right)^2} \frac{\wc}{\Ctot} \right] \\
        &\times \left[ -\frac{\Cj \Cgone \Cgone'}{\left(\Cgone + \Cj\right)^2} + \frac{\Cj \Cgone'}{\Cgone + \Cj} - \frac{\Cgone \Cgone'}{2 e^2} \frac{\partial^2 E_k}{\partial \Ng^2} + \frac{\Cgone^2 \Cgone'}{8 e^3} \Vg \frac{\partial^3 E_k}{\partial \Ng^3} \right]
\end{split}
\end{equation}
by plugging \eqref{eq:dwcdx} and \eqref{eq:dCeffdx_2} back to \eqref{eq:grp}.

Conversely, the cross-Kerr coupling arises from $\hbar \frac{1}{2} \frac{\partial^2 \wc}{\partial x^2} \hat{x}^2 \approx \hbar \frac{\partial^2 \wc}{\partial x^2} \xzp^2 b^\dagger b$ leading to
\begin{equation}
\label{eq:gck}
    \hbar \gck = \hbar \frac{\partial^2 \wc}{\partial x^2} \xzp^2 .
\end{equation}
A direct calculation leads to
\begin{equation}
\label{eq:d2wcdx2_appendix}
\begin{split}
        \frac{\partial^2 \wc}{\partial x^2} = & \wc \Cgtwo^2 \frac{\Cgtwo \left( 4\Cc + 3\Cgtwo \right) + 4\left( \Cc + \Cgtwo \right) \Ceff}{4 \left( \Cgtwo + \Ceff \right)^2 \left[ \Cc \Cgtwo + \left( \Cc + \Cgtwo \right) \Ceff \right]^2} \left( \frac{\partial \Ceff}{\partial x} \right)^2 \\
        &-\wc \Cgtwo^2 \frac{1}{2 \left( \Cgtwo + \Ceff \right) \left[ \Cc \Cgtwo + \left( \Cc + \Cgtwo \right) \Ceff \right]} \frac{\partial^2 \Ceff}{\partial x^2} ,
\end{split}
\end{equation}
where
\begin{equation}
\label{eq:d2Ceffdx2_appendix}
\begin{split}
    \frac{\partial^2 \Ceff}{\partial x^2} = & -\frac{\Cj \Cgone'^2}{\left( \Cgone + \Cj \right)^2} + \frac{\Cj \Cgone''}{\Cgone + \Cj} + \Cj \Cgone \left[ \frac{2 \Cgone'^2}{\left( \Cgone + \Cj \right)^3} - \frac{ \Cgone''}{\left( \Cgone + \Cj \right)^2} \right] \\
    &- \frac{1}{4 e^2} \left[ 4 \Cgone \Cgone' \frac{\partial}{\partial x} \left( \frac{\partial^2 E_k}{\partial \Ng^2} \right) + \left( 2 \Cgone'^2 + 2 \Cgone \Cgone'' \right) \frac{\partial^2 E_k}{\partial \Ng^2} + \Cgone^2 \frac{\partial^2}{\partial x^2} \left( \frac{\partial^2 E_k}{\partial \Ng^2} \right) \right] .
\end{split}
\end{equation}
Utilizing the second order chain rule $\frac{\partial^2 y}{\partial x^2} = \frac{\partial^2 y}{\partial z^2} \left( \frac{\partial z}{\partial x} \right)^2 + \frac{\partial y}{\partial z} \frac{\partial^2 z}{\partial x^2}$ we obtain
\begin{equation}
\label{eq:dx2_energy_band}
\begin{split}
    \frac{\partial^2}{\partial x^2} \left( \frac{\partial^2 E_k}{\partial \Ng^2} \right) &= \left( \frac{\partial \Ng}{\partial x} \right)^2 \frac{\partial^4 E_k}{\partial \Ng^4}  + \frac{\partial^2 \Ng}{\partial x^2} \frac{\partial^3 E_k}{\partial \Ng^3} \\
    &= \left( - \frac{\Vg}{2 e} \Cg' \right)^2 \frac{\partial^4 E_k}{\partial \Ng^4}  - \frac{\Vg}{2 e} \Cg'' \frac{\partial^3 E_k}{\partial \Ng^3}
\end{split}
\end{equation}
giving the explicit form of the second derivative
\begin{equation}
\label{eq:d2Ceffdx2_explicit_appendix}
\begin{split}
    \frac{\partial^2 \Ceff}{\partial x^2} = & -\frac{\Cj \Cgone'^2}{\left( \Cgone + \Cj \right)^2} + \frac{\Cj \Cgone''}{\Cgone + \Cj} + \Cj \Cgone \left[ \frac{2 \Cgone'^2}{\left( \Cgone + \Cj \right)^3} - \frac{ \Cgone''}{\left( \Cgone + \Cj \right)^2} \right] \\
    &- \frac{1}{4 e^2} \left[ \left( 2 \Cgone'^2 + 2 \Cgone \Cgone'' \right) \frac{\partial^2 E_k}{\partial \Ng^2} - \left( 2 \Cgone \Cgone'^2 \frac{\Vg}{e} + \Cgone^2 \Cg'' \frac{\Vg}{2 e} \right) \frac{\partial^3 E_k}{\partial \Ng^3} \right. \\
    & \ \ \ \ \ \ \ \ \ \ \left. + \Cgone^2 \left( - \frac{\Vg}{2 e} \Cg' \right)^2 \frac{\partial^4 E_k}{\partial \Ng^4} \right] .
\end{split}
\end{equation}
%

%XXXXXXXXXXXXXXXXXXXXXXXXXXXXXXXXXXXXXX
% QUANTIZING THE HAMILTONIAN
%XXXXXXXXXXXXXXXXXXXXXXXXXXXXXXXXXXXXXX
\section{Quantizing the Hamiltonian}
\label{sec:quantization}

Let us divide the full Hamiltonian \eqref{eq:total_Hamiltonian} into parts. The cavity, mechanics, cavity-mechanics-CPB coupling, charging, and Josephson-junction sub-Hamiltonians are
\begin{subequations}
\begin{align}
    \Hamil &= \Hc + \Hm + \Hcm + \Hch + \Hjj , \\
    \Hc &= \frac{1}{2 \Csigmac} Q_\mathrm{c}^2 + \frac{1}{2 \Lc} \phic^2 , \label{eq:cavity_Hamiltonian}\\
    \Hm &= 4 \Ec \left( \frac{\Csigmaonetwo^2}{\Csigmaone \Csigmatwo} - 1 \right) \Ng^2 , \label{eq:mechanics_Hamiltonian} \\
    \Hcm &=  \frac{\Csigmaonetwo}{\Csigmaone \Csigmatwoc} \cdot 2 e \Ng  Q_\mathrm{c} , \label{eq:c-m-coupling_Hamiltonian}\\
    \Hch &= 4 \Ec \left( n - \Ng \right)^2 + \frac{1}{\Csigmaonec} \cdot 2 e n Q_\mathrm{c} , \label{eq:charging_Hamiltonian_appendix}\\
    \Hjj &= - \Ej \cos \left( \pi \frac{\phie}{\Phi_0} \right) \left[ \cos \left( \varphi_1 \right) \cos \left( \frac{\varphi_\mathrm{c}}{2} \right) - \sin \left( \varphi_1 \right) \sin \left( \frac{\varphi_\mathrm{c}}{2}  \right) \right] \label{eq:JJ_Hamiltonian}\\
    &\ \ + \Ej d \sin \left( \pi \frac{\phie}{\Phi_0} \right) \left[ \cos \left( \varphi_1 \right) \sin \left( \frac{\varphi_\mathrm{c}}{2}  \right) + \sin \left( \varphi_1 \right) \cos \left( \frac{\varphi_\mathrm{c}}{2}  \right) \right] . \nonumber
\end{align}
\end{subequations}

%XXXXXXXXXXXXXXXXXXXXXXXXXXXXXXXXXXXXXX
% CAVITY HAMILTONIAN
%XXXXXXXXXXXXXXXXXXXXXXXXXXXXXXXXXXXXXX
\subsection{Cavity Hamiltonian}

Let us first quantize the cavity Hamiltonian \eqref{eq:cavity_Hamiltonian}. Quantized cavity flux and its conjugate momentum (charge) fulfill the canonical commutation relation
\begin{equation}
\label{eq:canonical_commutation}
    \left[ \hat{\phi}_\mathrm{c} , \hat{Q}_\mathrm{c} \right] = i \hbar ,
\end{equation}
and introduce bosonic operators $\hat{a},\hat{a}^\dagger$ with $\left[ \hat{a}, \hat{a}^\dagger \right] = 1$ so that
\begin{subequations}
\begin{align}
    \hat{\phi}_\mathrm{c} &= \phizp \left( \hat{a} + \hat{a}^\dagger \right) , \label{eq:cavity_flux_operator}\\
    \hat{Q}_\mathrm{c} &= -i \Qzp \left( \hat{a} - \hat{a}^\dagger \right) . \label{eq:cavity_charge_operator}
\end{align}
\end{subequations}
Applying these to the canonical commutation relation \eqref{eq:canonical_commutation} implies
\begin{equation}
\label{eq:phizp_qzp_relation}
    2 \phizp \Qzp = \hbar .
\end{equation}
Thus we can rewrite the quantized cavity Hamiltonian
\begin{equation}
    \hat{\Hamil}_\mathrm{c} = \left( \frac{\Qzp^2}{\Csigmac} + \frac{\phizp^2}{\Lc} \right) \left( \hat{a}^\dagger \hat{a} + \frac{1}{2} \right) + \left( - \frac{\Qzp^2}{2 \Csigmac} + \frac{\phizp^2}{2 \Lc} \right) \left( \hat{a}^2 + \hat{a}^{\dagger 2} \right) ,
\end{equation}
where we can denote $\hbar \wc = \frac{\Qzp^2}{\Csigmac} + \frac{\phizp^2}{\Lc}$. $\phizp$ and $\Qzp$ can be solved from this using the relation \eqref{eq:phizp_qzp_relation} and the standard way of writing the cavity angular frequency
\begin{equation}
\label{eq:cavity_freq}
    \wc = \frac{1}{\sqrt{\Lc \Csigmac}} .
\end{equation}
We find that
\begin{subequations}
\begin{align}
    \Qzp &= \sqrt{\frac{\hbar}{2 Z_0}} , \\
    \phizp &= \sqrt{\frac{\hbar Z_0}{2}} , \\
    Z_0 &= \sqrt{\frac{\Lc}{\Csigmac}} ,
\end{align}
\end{subequations}
and thus the cavity Hamiltonian can be written
\begin{equation}
\label{eq:q_cavity_Hamiltonian}
    \hat{\Hamil}_\mathrm{c} = \hbar \wc \left( \hat{a}^\dagger \hat{a} + \frac{1}{2} \right) ,
\end{equation}
where the constant term can be neglected without loss of generality.

%XXXXXXXXXXXXXXXXXXXXXXXXXXXXXXXXXXXXXX
% MECHANICS HAMILTONIAN
%XXXXXXXXXXXXXXXXXXXXXXXXXXXXXXXXXXXXXX
\subsection{Mechanics Hamiltonian}

The mechanics Hamiltonian \eqref{eq:mechanics_Hamiltonian} can be quantized in a similar fashion. Notice that the gate charge $\Ng$ is displacement $x$ dependent as the capacitance $\Cgone \left( x \right)$ depends on the  separation of the gate electrodes. We approximate
\begin{equation}
\label{eq:ng_approx}
    \Ng \approx \Ngzero + \frac{\partial \Ng}{\partial x} x
\end{equation}
and using the definition of $\Ng$ \eqref{eq:gate_charge_infinite_Cb_limit} we find
\begin{equation}
\label{eq:dngdx}
    \frac{\partial \Ng}{\partial x} = - \frac{\Cgone'}{2 e} \Vg .
\end{equation}
By defining bosonic operators for the mechanics $\hat{b},\hat{b}^\dagger$ with $\left[ \hat{b}, \hat{b}^\dagger \right] = 1$, we may quantize the displacement as
\begin{equation}
    \hat{x} = \xzp \left( \hat{b} + \hat{b}^\dagger \right) .
\end{equation}
The quantized mechanics Hamiltonian is thus (neglecting constants)
\begin{equation}
\label{eq:mechanics_Hamiltonian_q_appendix}
\begin{split}
    \hat{\Hamil}_\mathrm{m} &= 4 \Ec \left( \frac{\Csigmaonetwo^2}{\Csigmaone \Csigmatwo} - 1 \right) \left[ 2 \left( \frac{\partial \Ng}{\partial x} \right)^2 \xzp^2 \hat{b}^\dagger \hat{b} + 2 \Ng \frac{\partial \Ng}{\partial x} \xzp \left( \hat{b}^\dagger + \hat{b} \right) + \left( \frac{\partial \Ng}{\partial x} \right)^2 \xzp^2 \left( \hat{b}^{\dagger 2} + \hat{b}^2 \right) \right] \\
    &= \hbar \wm \hat{b}^\dagger \hat{b} + h_1 \left( \hat{b}^\dagger + \hat{b} \right) + h_2 \left( \hat{b}^{\dagger 2} + \hat{b}^2 \right) .
\end{split}
\end{equation}
We now have enough information to also quantize $\Hcm$ \eqref{eq:c-m-coupling_Hamiltonian} that directly couples the cavity to the mechanics, and obtain
\begin{equation}
\label{eq:c-m-coupling_Hamiltonian_q}
\begin{split}
    \hat{\mathcal{H}}_\mathrm{cm} &= \frac{\Csigmaonetwo}{\Csigmaone \Csigmatwoc} \cdot 2 e \left( \Ngzero + \frac{\partial \Ng}{\partial x} \hat{x} \right)  \hat{Q}_\mathrm{c} \\
    &= -i 2 e \frac{\Csigmaonetwo}{\Csigmaone \Csigmatwoc} \left[ \Ngzero - \frac{\Cgone'}{2 e} \Vg \xzp \left( \hat{b} + \hat{b}^\dagger \right) \right]  \Qzp \left( \hat{a} - \hat{a}^\dagger \right) .
\end{split}
\end{equation}

%XXXXXXXXXXXXXXXXXXXXXXXXXXXXXXXXXXXXXX
% JJ HAMILTONIAN
%XXXXXXXXXXXXXXXXXXXXXXXXXXXXXXXXXXXXXX
\subsection{Josephson junction Hamiltonian}

In order to quantize the Josephson junction Hamiltonian \eqref{eq:JJ_Hamiltonian}, let us first discuss how the phase on the CPB island relates to the qubit operations when the tunneling junctions are considered as two-level systems. The following derivation closely follows the treatment of tunnel junctions by Vool and Devoret\cite{bib:Vool2017}. The tunnelling Hamiltonian of a Josephson junction can be written in a number basis using the transmitted charge through the junction, which in terms of Cooper pairs reads as
\begin{equation}
\label{eq:tunneling_Hamiltonian}
    \hat{\Hamil}_\mathrm{T} = - \frac{E_\mathrm{T}}{2} \sum_{N=-\infty}^{N=\infty} \Big[ \ket{N}\bra{N+1} + \ket{N+1}\bra{N} \Big],
\end{equation}
where the tunneling energy is denoted by $E_\mathrm{T}$. This number basis representation can be related to the alternate phase basis by equations
\begin{subequations}
\begin{align}
    \ket{\varphi} &= \sum_{N=-\infty}^{\infty} e^{iN\varphi} \ket{N} , \\
    \ket{N} &= \frac{1}{2\pi} \int_0^{2\pi} d\varphi e^{-iN\varphi} \ket{\varphi} .
\end{align}
\end{subequations}
A straightforward calculation reveals that the operator
\begin{equation}
    e^{i\hat{\varphi}} = \frac{1}{2\pi} \int_0^{2\pi} d\phi e^{i\varphi} \ket{\varphi}\bra{\varphi}
\end{equation}
has the following translation properties
\begin{subequations}
\begin{align}
    e^{i\hat{\varphi}} \ket{N} &= \ket{N-1} , \\
    e^{-i\hat{\varphi}} \ket{N} &= \ket{N+1} .
\end{align}
\end{subequations}
Confining our approach to a two level setting, i.e. we only consider qubit states $\ket{0}$ and $\ket{1}$, directly leads to
\begin{subequations}
\begin{align}
    \cos\hat{\varphi} &= \frac{1}{2} \left[ e^{i\hat{\varphi}} + e^{-i\hat{\varphi}} \right] = \frac{1}{2} \Big[\ket{1}\bra{0} + \ket{0}\bra{1}\Big] = \frac{1}{2} \sigma_x , \\
    \sin\hat{\varphi} &= \frac{1}{2i} \left[ e^{i\hat{\varphi}} - e^{-i\hat{\varphi}} \right] = \frac{i}{2} \Big[\ket{1}\bra{0} - \ket{0}\bra{1}\Big] = -\frac{1}{2} \sigma_y .
\end{align}
\end{subequations}
We may thus identify
\begin{subequations}
\begin{align}
    \cos \left( \hat{\varphi}_1 \right) &= \frac{1}{2} \sigma_x , \\
    \sin \left( \hat{\varphi}_1 \right) &= - \frac{1}{2} \sigma_y ,
\end{align}
\end{subequations}
where $\sigma_k$ are the Pauli matrices with conventions
\begin{subequations}
\begin{align}
    \sigma_x &= \ket{1}\bra{0} + \ket{0}\bra{1} = \begin{pmatrix} 0 & 1\\ 1 & 0 \end{pmatrix}, \\
    \sigma_y &= i \left( \ket{0}\bra{1} - \ket{1}\bra{0} \right) = i \begin{pmatrix} 0 & -1\\ 1 & 0 \end{pmatrix}, \\
    \ket{0} &= \begin{pmatrix} 0 \\ 1 \end{pmatrix}, \\
    \ket{1} &= \begin{pmatrix} 1 \\ 0 \end{pmatrix}.
\end{align}
\end{subequations}
By recall the cavity flux operator \eqref{eq:cavity_flux_operator} and the phase-flux relation $\varphi_\mathrm{c} = 2 \pi \frac{\phi_c}{\Phi_0}$, we can  write
\begin{subequations}
\begin{align}
    \frac{\hat{\varphi}_\mathrm{c}}{2} &= \eta \left( \hat{a} + \hat{a}^\dagger \right) , \\
    \eta &= \sqrt{\frac{e^2 Z_0}{2 \hbar}} .
\end{align}
\end{subequations}
We can expand the trigonometric functions of the cavity flux operators in the JJ Hamiltonian \eqref{eq:JJ_Hamiltonian}
\begin{subequations}
\begin{align}
    \sin \left( \frac{\hat{\varphi}_\mathrm{c}}{2} \right) &\approx \frac{\hat{\varphi}_\mathrm{c}}{2} = \eta \left( \hat{a} + \hat{a}^\dagger \right) , \\
    \cos \left( \frac{\hat{\varphi}_\mathrm{c}}{2} \right) &\approx 1- \frac{1}{2} \left( \frac{\hat{\varphi}_\mathrm{c}}{2} \right)^2 = 1 - \frac{1}{2} \eta^2 \left( \hat{a} + \hat{a}^\dagger \right)^2 ,
\end{align}
\end{subequations}
and the Josephson junction Hamiltonian simplifies to
\begin{equation}
\label{eq:JJ_Hamiltonian_q}
\begin{split}
    \hat{\Hamil}_\mathrm{JJ} =& - \frac{\Ej}{2} \left[ \cos \left( \pi \frac{\phie}{\Phi_0} \right) \sigma_x + d \sin \left( \pi \frac{\phie}{\Phi_0} \right) \sigma_y \right] \\
    & - \frac{\Ej}{2} \left[ \cos \left( \pi \frac{\phie}{\Phi_0} \right) \sigma_y - d \sin \left( \pi \frac{\phie}{\Phi_0} \right) \sigma_x \right] \eta \left( \hat{a} + \hat{a}^\dagger \right)\\
    &- \frac{\Ej}{4} \left[ \cos \left( \pi \frac{\phie}{\Phi_0} \right) \sigma_x + d \sin \left( \pi \frac{\phie}{\Phi_0} \right) \sigma_y \right] \eta^2 \left( \hat{a} + \hat{a}^\dagger \right)^2 \\
    =& - \frac{B_1}{2} \sigma_x - \frac{B_2}{2} \sigma_y + g_1 \sigma_y \xc + g_2 \sigma_x \xc^2 + g_3 \sigma_x \xc  + g_4 \sigma_y \xc^2
\end{split}
\end{equation}
with
\begin{subequations}
\begin{align}
    B_1 &= \Ej \cos \left( \pi \frac{\phie}{\Phi_0} \right) , \\
    B_2 &= \Ej d \sin \left( \pi \frac{\phie}{\Phi_0} \right) , \\
    g_1 &= -\frac{B_1}{2} \eta , \\
    g_2 &= \frac{B_1}{4} \eta^2 , \\
    g_3 &= \frac{B_2}{2} \eta , \\
    g_4 &= \frac{B_2}{4} \eta^2 , \\
    \xc &= \hat{a} + \hat{a}^\dagger .
\end{align}
\end{subequations}

%XXXXXXXXXXXXXXXXXXXXXXXXXXXXXXXXXXXXXX
% CHARGING HAMILTONIAN
%XXXXXXXXXXXXXXXXXXXXXXXXXXXXXXXXXXXXXX
\subsection{Charging Hamiltonian}

The charging Hamiltonian \eqref{eq:charging_Hamiltonian_appendix} can be written in a quantized form
\begin{equation}
\label{eq:charging_Hamiltonian_q_appendix}
    \hat{\Hamil}_\mathrm{ch} = 4 \Ec \sum_n \left[ \left( \hat{n} - \Ng \right)^2 + \frac{2e}{\Csigmaonec} \hat{n} \hat{Q}_\mathrm{c} \right] \ket{n}\bra{n}
\end{equation}
with eigenenergies $E_n = 4 \Ec \left( n - \Ng \right)^2 + \frac{2e}{\Csigmaonec} n Q_\mathrm{c}$ that are parabolas as the function of $\Ng$ displaced by the cavity charge. Assuming that the $\Ej \ll \Ec$, the contribution from the JJ Hamiltonian \eqref{eq:JJ_Hamiltonian_q} to the qubit energy is negligible and the total qubit energy can be approximated with the parabolas $E_n$.

Concentrate on the states closest to $\Ngzero$, i.e. $n = \mathrm{int} \left( \Ngzero \right)$ and $n = \mathrm{int} \left( \Ngzero \right) + 1$, and call these states $\ket{0}$ and $\ket{1}$, respectively. From $E_n$, we can see that the degeneracy point of this well-defined qubit is at $\Ng = \frac{1}{2} + \frac{1}{8 \Ec} \frac{2e}{\Csigmaonec} Q_\mathrm{c}$.

Thus, in the two-level system approximation, the charging Hamiltonian is
\begin{equation}
\begin{split}
    \hat{\Hamil}_\mathrm{ch} &= 2 \Ec \left[ 1 - \left( \frac{1}{2} + \frac{1}{8 \Ec} \frac{2e}{\Csigmaonec} \hat{Q}_\mathrm{c} \right)^{-1} \Ng \right] \sigma_z \\
    &= 2 \Ec \left[ 1 - \left( \frac{1}{2} + \frac{1}{8 \Ec} \frac{2e}{\Csigmaonec} \hat{Q}_\mathrm{c} \right)^{-1} \Ngzero \right] \sigma_z - 2 \Ec \left( \frac{1}{2} + \frac{1}{8 \Ec} \frac{2e}{\Csigmaonec} \hat{Q}_\mathrm{c} \right)^{-1} \frac{\partial \Ng}{\partial x} \hat{x} \sigma_z .
\end{split}
\end{equation}
We can expand $\left( \frac{1}{2} + \frac{1}{8 \Ec} \frac{2e}{\Csigmaonec} Q_\mathrm{c} \right)^{-1} \approx 2 - \frac{1}{\Ec} \frac{e}{\Csigmaonec} \hat{Q}_\mathrm{c}$ which allows us to write the charging Hamiltonian
\begin{equation}
\begin{split}
    \hat{\Hamil}_\mathrm{ch} &= 2 \Ec \left( 1 - 2 \Ngzero \right) \sigma_z - 2 \frac{e}{\Csigmaonec} \Qzp \Ngzero \pc \sigma_z \\
    & \ \ - 4 \Ec \frac{\partial \Ng}{\partial x} \xzp \xm \sigma_z + 2 \frac{e}{\Csigmaonec} \frac{\partial \Ng}{\partial x} \xzp \Qzp \xm \pc \sigma_z \\
    &= - \frac{B_3}{2} \sigma_z - \gcp \pc \sigma_z - \gm \xm \sigma_z + \gcm \xm \pc \sigma_z ,
\end{split}
\end{equation}
where
\begin{subequations}
\begin{align}
    B_3 &= -4 \Ec \left( 1 - 2 \Ngzero \right) , \\
    \gm &= 4 \Ec \frac{\partial \Ng}{\partial x} \xzp = 4 \Ec \left( - \frac{1}{2e} \Cgone' \Vg \right) \xzp = - \frac{2}{e} \Ec \Vg \Cgone' \xzp , \\
    \gcp &= 2 \frac{e}{\Csigmaonec} \Qzp \Ngzero , \\
    \gcm &= 2 \frac{e}{\Csigmaonec} \frac{\partial \Ng}{\partial x} \Qzp \xzp , \\
    \xm &= \hat{b} + \hat{b}^\dagger , \\
    \pc &= -i \left( \hat{a} - \hat{a}^\dagger \right) .
\end{align}
\end{subequations}

%XXXXXXXXXXXXXXXXXXXXXXXXXXXXXXXXXXXXXX
% PERTURBATIVE APPROACH TO THE QUBIT
%XXXXXXXXXXXXXXXXXXXXXXXXXXXXXXXXXXXXXX
\section{Perturbative approach to the qubit}
\label{sec:appendix_QM_perturbation}

Let us regroup the quantized Hamiltonians based on their effects on the qubit
\begin{subequations}
\begin{align}
    \hat{\Hamil} &= \hat{\Hamil}_0 + \hat{\Hamil}_x + \hat{\Hamil}_y + \hat{\Hamil}_z ,\\
    \hat{\Hamil}_0 &= \hat{\Hamil}_\mathrm{c} + \hat{\Hamil}_\mathrm{m} + \hat{\Hamil}_\mathrm{cm}, \\
    \hat{\Hamil}_x &= \left[ - \frac{1}{2} B_1 + g_3 \xc + g_2 \xc^2 \right] \sigma_x = - \frac{1}{2} \tilde{B}_1 \sigma_x, \\
    \hat{\Hamil}_y &= \left[ - \frac{1}{2} B_2 + g_1 \xc + g_4 \xc^2 \right] \sigma_y = - \frac{1}{2} \tilde{B}_2 \sigma_y, \\
    \hat{\Hamil}_z &= \left[ - \frac{1}{2} B_3 - \gm \xm - \gcp \pc + \gcm \pc \xm \right] \sigma_z = - \frac{1}{2} \tilde{B}_3 \sigma_z
    %\hat{\Hamil}_z &= \left[ - \frac{1}{2} B_3 - \gm \xm  \right] \sigma_z = - \frac{1}{2} \tilde{B}_3 \sigma_z 
\end{align}
\end{subequations}
with
\begin{subequations}
\begin{align}
    \tilde{B}_1 &= B_1 - 2 g_3 \xc - 2 g_2 \xc^2 , \\
    \tilde{B}_2 &= B_2 - 2 g_1 \xc - 2 g_4 \xc^2 , \\
    \tilde{B}_3 &= B_3 + 2 \gm \xm + 2 \gcp \pc - 2 \gcm \pc \xm.
\end{align}
\end{subequations}
Provided that $\Ec \gg \Ej$, the additional terms in $\tilde{B}_j$ on can be considered as a small perturbation to the unperturbed qubit Hamiltonian
\begin{equation}
    \hat{\mathcal{H}}_{\mathrm{Q}0} = - \frac{1}{2} \sum_j B_j \sigma_j .
\end{equation}
We can treat our system with a perturbation approach to the $\hat{H}_{\mathrm{Q}0}$ with eigenenergies $\pm \frac{1}{2} \sqrt{B_1^2 + B_2^2 + B_3^2}$. Thus we can approximate the full Hamiltonian as a perturbation to the qubit ground state
\begin{equation}
\label{eq:perturbation_Hamiltonian}
    \hat{\mathcal{H}} = \hat{\Hamil}_0 - \frac{1}{2} \sqrt{\tilde{B}_1^2 + \tilde{B}_2^2 + \tilde{B}_3^2} .
\end{equation}
Let us write in the above expression explicitly
\begin{equation}
\begin{split}
    & - \frac{1}{2} \sqrt{\tilde{B}_1^2 + \tilde{B}_2^2 + \tilde{B}_3^2} \\
    =& -\frac{1}{2} B \sqrt{1 + \frac{1}{B^2} \left( \alpha \xc + \beta \xc^2 + \rho \xc^3 + \delta \xc^4 + \epsilon \xm + \lambda \xm^2 + \xi_1 \pc + \xi_2 \pc^2 + \xi_3 \pc \xm + \xi_4 \pc^2 \xm + \xi_5 \pc \xm^2 \right)}
\end{split}
\end{equation}
with
\begin{subequations}
\begin{align}
    B &= \sqrt{B_1^2 + B_2^2 + B_3^2} , \\
    \alpha &= - 4 \left( B_1 g_3 + B_2 g_1 \right) , \\
    \beta &= 4 \left( g_3^2 + g_1^2 - B_1 g_2 - B_2 g_4 \right) , \\
    \rho &= 8 \left( g_2 g_3  + g_1 g_4 \right) , \\
    \delta &= 4 \left( g_2^2 + g_4^2 \right) , \\
    \epsilon &= 4 B_3 \gm , \\
    \lambda &= 4 \gm^2 , \\
    \xi_1 &= 4 B_3 \gcp , \\
    \xi_2 &= 4 \gcp^2 , \\
    \xi_3 &= 4 \left( 2 \gcp \gm - B_3 \gcm \right) , \\
    \xi_4 &= -8 \gcp \gcm , \\
    \xi_5 &= - 8 \gm \gcm .
\end{align}
\end{subequations}

\begin{figure}[tbh]
    \centering
    \begin{subfigure}[]{0.47\textwidth}
        \includegraphics[width=\textwidth]{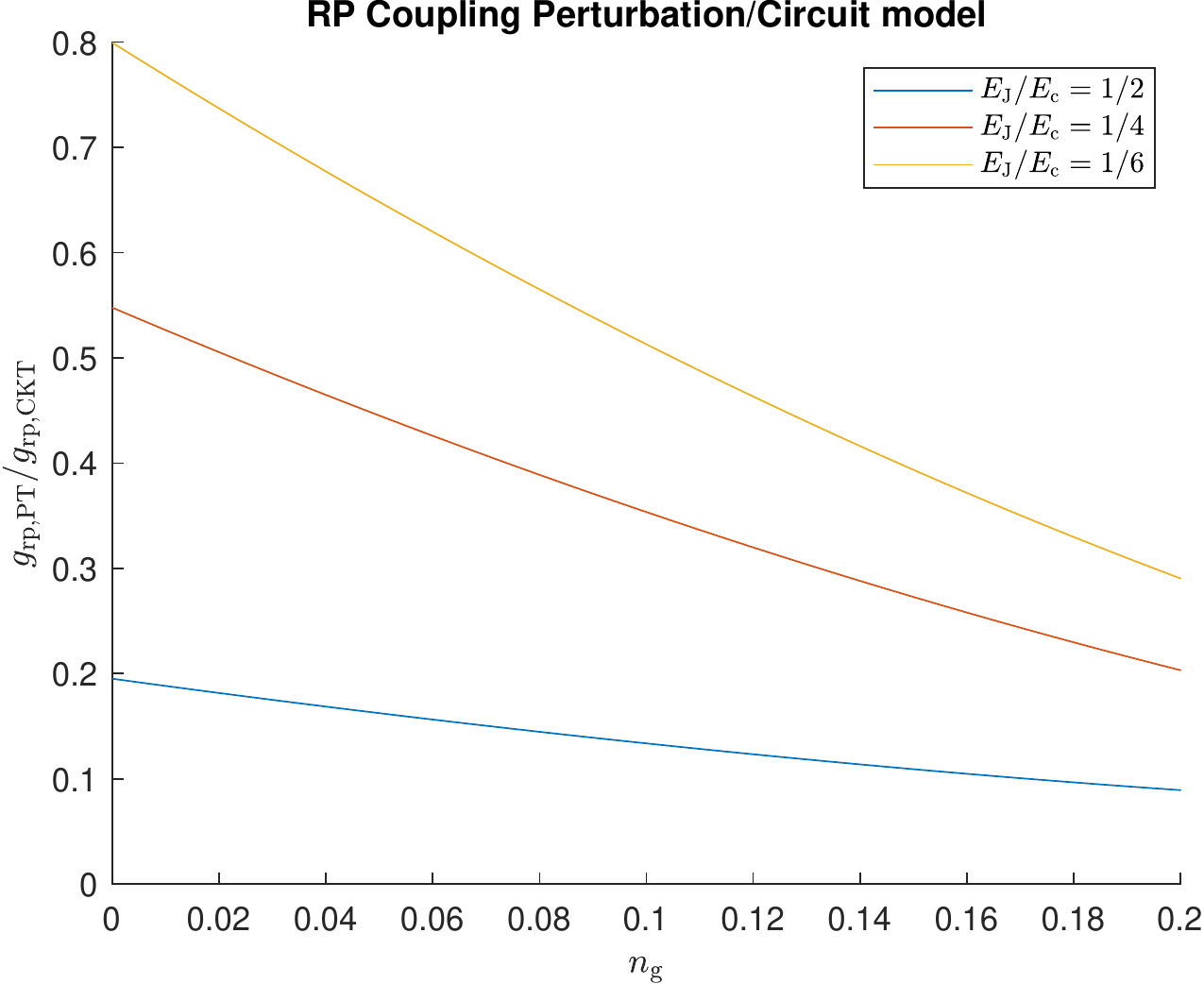}
        \caption{}
        \label{fig:grp_comparison}
    \end{subfigure}
    \begin{subfigure}[]{0.47\textwidth}
        \includegraphics[width=\textwidth]{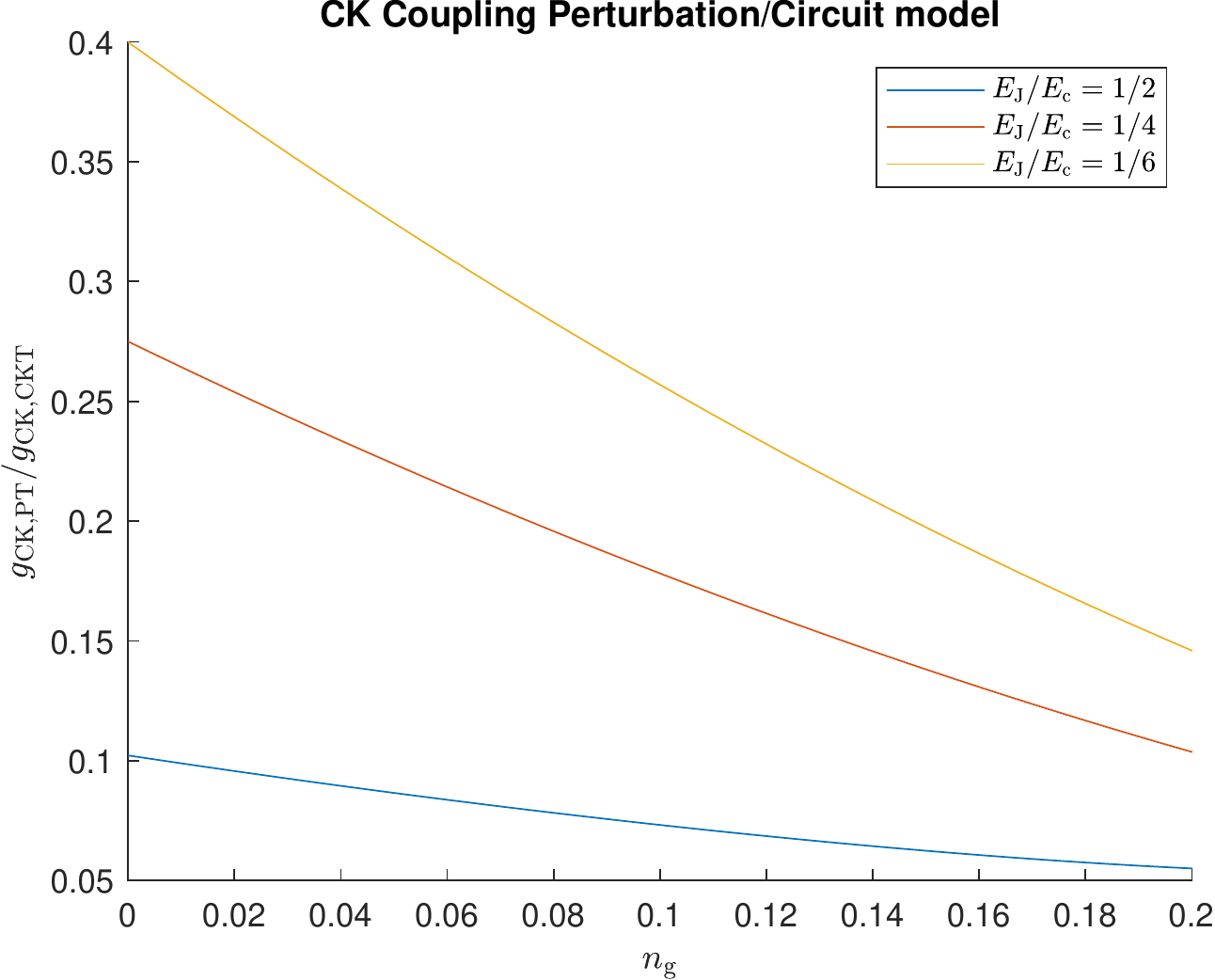}
        \caption{}
        \label{fig:gck_comparison}
    \end{subfigure}
    \label{fig:coupling_comparisons}
    \caption{\textbf{(a)} Comparison between radiation pressure coupling from the quantum mechanical perturbation theory and the circuit model. Ratio between them approaches 1 far away from charge degeneracy points as $\Ej/\Ec$ ratio becomes smaller. \textbf{(b)} Similar comparison for the cross-Kerr couplings obtained from the two approaches.}
\end{figure}

Recall that $\sqrt{1+x} \approx 1 + \frac{x}{2} - \frac{x^2}{8} + \frac{x^3}{16}$ which allows us to expand the above expression provided that $B$ is large, i.e. $\Ec$ is large
\begin{equation}
\label{eq:sqrt_expansion_appendix}
\begin{split}
    & - \frac{1}{2} \sqrt{\tilde{B}_1^2 + \tilde{B}_2^2 + \tilde{B}_3^2} \\
    \approx & - \frac{1}{2} B \\
    & - \frac{1}{4 B} \left( \alpha \xc + \beta \xc^2 + \rho \xc^3 + \delta \xc^4 + \epsilon \xm + \lambda \xm^2 \right. \\
    & \ \ \ \ \ \ \ \ \ \left. + \xi_1 \pc + \xi_2 \pc^2 + \xi_3 \pc \xm + \xi_4 \pc^2 \xm + \xi_5 \pc \xm^2 \right) \\
    & + \frac{1}{16 B^3} \left( \alpha \xc + \beta \xc^2 + \rho \xc^3 + \delta \xc^4 + \epsilon \xm + \lambda \xm^2 \right. \\
    & \ \ \ \ \ \ \ \ \ \left. + \xi_1 \pc + \xi_2 \pc^2 + \xi_3 \pc \xm + \xi_4 \pc^2 \xm + \xi_5 \pc \xm^2 \right)^2 .
\end{split}
\end{equation}
We can now identify the terms contributing the radiation pressure coupling $\left( \xc^{2k} \pc^{2j} \xm^{2l+1} \right)$, cross-Kerr coupling $\left( \xc^{2k} \pc^{2j} \xm^{2l} \right)$. We also need to take into account the prefactors arising from normal ordering the terms contributing to these couplings $\hat{a}^\dagger \hat{a} \left( \hat{b}^\dagger + \hat{b} \right)$, $\hat{a}^\dagger \hat{a} \hat{b}^\dagger \hat{b}$, respectively.

The radiation pressure coupling is
\begin{equation}
\label{eq:SWT_grp_appendix}
\begin{split}
    -\hbar \grp = - \frac{1}{2 B} \xi_4 + \frac{1}{4 B^3} \left[ \epsilon \left( \beta + 6 \delta + \xi_2 \right) + \xi_4 \left( 2 \beta - 3 \delta + 3 \lambda \right) \right]
\end{split}
\end{equation}
(note the minus sign fixing the radiation pressure coupling to $\grp = -\frac{\partial \wc}{\partial x}$ , i.e. $\hbar \wc \left(x\right) \hat{a}^\dagger \hat{a} \approx \hbar \left( \wc - \grp \xm \right) \hat{a}^\dagger \hat{a}$) and the cross-Kerr coupling is
\begin{equation}
\label{eq:SWT_gck_appendix}
\begin{split}
    \hbar \gck = \frac{1}{4 B^3} \left[ 2 \lambda \left( \beta + 6 \delta \right) + \xi_3^2 + 6 \xi_4^2 + 6 \xi_5^2 + 2 \epsilon \xi_4 + 2 \lambda \xi_4 \right] .
\end{split}
\end{equation}
Ignoring the asymmetry in the qubit eigenenergies introduced by the cavity charge in Eq. \eqref{eq:charging_Hamiltonian_q_appendix}, we can more easily expand Eq. \eqref{eq:sqrt_expansion_appendix} to higher order. These terms are accompanied with $\pc$ operators that do not commute with $\xc$ thus making high order calculations significantly more cumbersome. Omitting these terms is valid, since the terms $\xi_i$ resulting from this cavity charge induced asymmetry are orders of magnitude smaller compared to the other terms in the expansion. In this approximation, the radiation pressure and cross-Kerr couplings are in the third order expansion
\begin{subequations}
\begin{align}
    \hbar \grp =& \frac{\epsilon}{16 B^5} \left\{ - 4 B^2 \left( \beta + 6 \delta \right) + 3 \alpha^2 + 36 \alpha \rho + 135 \rho^2 + 18 \left( \beta^2 + 15 \beta \delta + 70 \delta^2 + \beta \lambda + 6 \delta \lambda \right) \right\} , \label{eq:SWT_grp_3rd_appendix} \\
    \hbar \gck =& - \frac{1}{8 B^5} \left\{ \lambda \left[ -4 B^2 \left( \beta + 6 \delta \right) + 3 \alpha^2 + 36 \alpha \rho + 135 \rho^2 + 18 \left( \beta^2 + 15 \beta \delta + 70 \delta^2 \right) \right]  \right. \label{eq:SWT_gck_3rd_appendix} \\
    & \left. + 3 \left( \beta + 6 \delta \right) \epsilon^2 + 18 \left( \beta + 6 \delta \right) \lambda^2 \right\} . \nonumber
\end{align}
\end{subequations}

The third order expansion for the radiation pressure coupling is not crucial for the result to align well with the circuit model far away from the degeneracy point of the qubit. However, for the cross-Kerr coupling, expanding to the third order offers a much better agreement with the circuit model than just the second order expansion.

In Fig. \ref{fig:grp_comparison}, we see that as $\Ej \ll \Ec$ limit is approached, the radiation pressure coupling arising from the perturbative quantum mechanical method approaches the result from the circuit model far away from the degeneracy point of the qubit. Similarly, in Fig. \ref{fig:gck_comparison}, results for the cross-Kerr couplings from the two approaches become better aligned as $\Ej/\Ec$ ratio decreases. Here, for the values obtained from the quantum mechanical approach, the second order result \eqref{eq:SWT_grp_appendix} is used for the radiation pressure coupling while the cross-Kerr comparison is calculated with the third order equation \eqref{eq:SWT_gck_3rd_appendix}.

%xxxxxxxxxxxxxxxxxxxxxxxxxxxxxxxxxxxxxxxxxxx
% THE BIBLIOGRAPHY
%xxxxxxxxxxxxxxxxxxxxxxxxxxxxxxxxxxxxxxxxxxx

\twocolumngrid
\bibliography{cset_bib2}

\begin{thebibliography}{41}%
\makeatletter
\providecommand \@ifxundefined [1]{%
 \@ifx{#1\undefined}
}%
\providecommand \@ifnum [1]{%
 \ifnum #1\expandafter \@firstoftwo
 \else \expandafter \@secondoftwo
 \fi
}%
\providecommand \@ifx [1]{%
 \ifx #1\expandafter \@firstoftwo
 \else \expandafter \@secondoftwo
 \fi
}%
\providecommand \natexlab [1]{#1}%
\providecommand \enquote  [1]{``#1''}%
\providecommand \bibnamefont  [1]{#1}%
\providecommand \bibfnamefont [1]{#1}%
\providecommand \citenamefont [1]{#1}%
\providecommand \href@noop [0]{\@secondoftwo}%
\providecommand \href [0]{\begingroup \@sanitize@url \@href}%
\providecommand \@href[1]{\@@startlink{#1}\@@href}%
\providecommand \@@href[1]{\endgroup#1\@@endlink}%
\providecommand \@sanitize@url [0]{\catcode `\\12\catcode `\$12\catcode
  `\&12\catcode `\#12\catcode `\^12\catcode `\_12\catcode `\%12\relax}%
\providecommand \@@startlink[1]{}%
\providecommand \@@endlink[0]{}%
\providecommand \url  [0]{\begingroup\@sanitize@url \@url }%
\providecommand \@url [1]{\endgroup\@href {#1}{\urlprefix }}%
\providecommand \urlprefix  [0]{URL }%
\providecommand \Eprint [0]{\href }%
\providecommand \doibase [0]{http://dx.doi.org/}%
\providecommand \selectlanguage [0]{\@gobble}%
\providecommand \bibinfo  [0]{\@secondoftwo}%
\providecommand \bibfield  [0]{\@secondoftwo}%
\providecommand \translation [1]{[#1]}%
\providecommand \BibitemOpen [0]{}%
\providecommand \bibitemStop [0]{}%
\providecommand \bibitemNoStop [0]{.\EOS\space}%
\providecommand \EOS [0]{\spacefactor3000\relax}%
\providecommand \BibitemShut  [1]{\csname bibitem#1\endcsname}%
\let\auto@bib@innerbib\@empty
%</preamble>
\bibitem [{\citenamefont {Aspelmeyer}\ \emph {et~al.}(2014)\citenamefont
  {Aspelmeyer}, \citenamefont {Kippenberg},\ and\ \citenamefont
  {Marquardt}}]{bib:Aspelmeyer2014}%
  \BibitemOpen
  \bibfield  {author} {\bibinfo {author} {\bibfnamefont {M.}~\bibnamefont
  {Aspelmeyer}}, \bibinfo {author} {\bibfnamefont {T.~J.}\ \bibnamefont
  {Kippenberg}}, \ and\ \bibinfo {author} {\bibfnamefont {F.}~\bibnamefont
  {Marquardt}},\ }\href {\doibase 10.1103/RevModPhys.86.1391} {\bibfield
  {journal} {\bibinfo  {journal} {Reviews of Modern Physics}\ }\textbf
  {\bibinfo {volume} {86}},\ \bibinfo {pages} {1391} (\bibinfo {year}
  {2014})}\BibitemShut {NoStop}%
\bibitem [{\citenamefont {Armour}\ \emph {et~al.}(2002)\citenamefont {Armour},
  \citenamefont {Blencowe},\ and\ \citenamefont {Schwab}}]{bib:Armour2002}%
  \BibitemOpen
  \bibfield  {author} {\bibinfo {author} {\bibfnamefont {A.~D.}\ \bibnamefont
  {Armour}}, \bibinfo {author} {\bibfnamefont {M.~P.}\ \bibnamefont
  {Blencowe}}, \ and\ \bibinfo {author} {\bibfnamefont {K.~C.}\ \bibnamefont
  {Schwab}},\ }\href {\doibase 10.1103/PhysRevLett.88.148301} {\bibfield
  {journal} {\bibinfo  {journal} {Physical Review Letters}\ }\textbf {\bibinfo
  {volume} {88}},\ \bibinfo {pages} {148301} (\bibinfo {year}
  {2002})}\BibitemShut {NoStop}%
\bibitem [{\citenamefont {Regal}\ \emph {et~al.}(2008)\citenamefont {Regal},
  \citenamefont {Teufel},\ and\ \citenamefont {Lehnert}}]{bib:Regal2008}%
  \BibitemOpen
  \bibfield  {author} {\bibinfo {author} {\bibfnamefont {C.~A.}\ \bibnamefont
  {Regal}}, \bibinfo {author} {\bibfnamefont {J.~D.}\ \bibnamefont {Teufel}}, \
  and\ \bibinfo {author} {\bibfnamefont {K.~W.}\ \bibnamefont {Lehnert}},\
  }\href {http://dx.doi.org/10.1038/nphys974} {\bibfield  {journal} {\bibinfo
  {journal} {Nature Physics}\ }\textbf {\bibinfo {volume} {4}},\ \bibinfo
  {pages} {555} (\bibinfo {year} {2008})}\BibitemShut {NoStop}%
\bibitem [{\citenamefont {LaHaye}\ \emph {et~al.}(2009)\citenamefont {LaHaye},
  \citenamefont {Suh}, \citenamefont {Echternach}, \citenamefont {Schwab},\
  and\ \citenamefont {Roukes}}]{bib:LaHaye2009}%
  \BibitemOpen
  \bibfield  {author} {\bibinfo {author} {\bibfnamefont {M.~D.}\ \bibnamefont
  {LaHaye}}, \bibinfo {author} {\bibfnamefont {J.}~\bibnamefont {Suh}},
  \bibinfo {author} {\bibfnamefont {P.~M.}\ \bibnamefont {Echternach}},
  \bibinfo {author} {\bibfnamefont {K.~C.}\ \bibnamefont {Schwab}}, \ and\
  \bibinfo {author} {\bibfnamefont {M.~L.}\ \bibnamefont {Roukes}},\ }\href
  {http://dx.doi.org/10.1038/nature08093} {\bibfield  {journal} {\bibinfo
  {journal} {Nature}\ }\textbf {\bibinfo {volume} {459}},\ \bibinfo {pages}
  {960} (\bibinfo {year} {2009})}\BibitemShut {NoStop}%
\bibitem [{\citenamefont {Rocheleau}\ \emph {et~al.}(2010)\citenamefont
  {Rocheleau}, \citenamefont {Ndukum}, \citenamefont {Macklin}, \citenamefont
  {Hertzberg}, \citenamefont {Clerk},\ and\ \citenamefont
  {Schwab}}]{bib:Rochelau2010}%
  \BibitemOpen
  \bibfield  {author} {\bibinfo {author} {\bibfnamefont {T.}~\bibnamefont
  {Rocheleau}}, \bibinfo {author} {\bibfnamefont {T.}~\bibnamefont {Ndukum}},
  \bibinfo {author} {\bibfnamefont {C.}~\bibnamefont {Macklin}}, \bibinfo
  {author} {\bibfnamefont {J.~B.}\ \bibnamefont {Hertzberg}}, \bibinfo {author}
  {\bibfnamefont {A.~A.}\ \bibnamefont {Clerk}}, \ and\ \bibinfo {author}
  {\bibfnamefont {K.~C.}\ \bibnamefont {Schwab}},\ }\href
  {http://dx.doi.org/10.1038/nature08681} {\bibfield  {journal} {\bibinfo
  {journal} {Nature}\ }\textbf {\bibinfo {volume} {463}},\ \bibinfo {pages}
  {72} (\bibinfo {year} {2010})}\BibitemShut {NoStop}%
\bibitem [{\citenamefont {Teufel}\ \emph {et~al.}(2011)\citenamefont {Teufel},
  \citenamefont {Donner}, \citenamefont {Li}, \citenamefont {Harlow},
  \citenamefont {Allman}, \citenamefont {Cicak}, \citenamefont {Sirois},
  \citenamefont {Whittaker}, \citenamefont {Lehnert},\ and\ \citenamefont
  {Simmonds}}]{bib:Teufel2011}%
  \BibitemOpen
  \bibfield  {author} {\bibinfo {author} {\bibfnamefont {J.~D.}\ \bibnamefont
  {Teufel}}, \bibinfo {author} {\bibfnamefont {T.}~\bibnamefont {Donner}},
  \bibinfo {author} {\bibfnamefont {D.}~\bibnamefont {Li}}, \bibinfo {author}
  {\bibfnamefont {J.~W.}\ \bibnamefont {Harlow}}, \bibinfo {author}
  {\bibfnamefont {M.~S.}\ \bibnamefont {Allman}}, \bibinfo {author}
  {\bibfnamefont {K.}~\bibnamefont {Cicak}}, \bibinfo {author} {\bibfnamefont
  {A.~J.}\ \bibnamefont {Sirois}}, \bibinfo {author} {\bibfnamefont {J.~D.}\
  \bibnamefont {Whittaker}}, \bibinfo {author} {\bibfnamefont {K.~W.}\
  \bibnamefont {Lehnert}}, \ and\ \bibinfo {author} {\bibfnamefont {R.~W.}\
  \bibnamefont {Simmonds}},\ }\href {http://dx.doi.org/10.1038/nature10261}
  {\bibfield  {journal} {\bibinfo  {journal} {Nature}\ }\textbf {\bibinfo
  {volume} {475}},\ \bibinfo {pages} {359} (\bibinfo {year}
  {2011})}\BibitemShut {NoStop}%
\bibitem [{\citenamefont {Nation}\ \emph {et~al.}(2016)\citenamefont {Nation},
  \citenamefont {Suh},\ and\ \citenamefont {Blencowe}}]{bib:Nation2016}%
  \BibitemOpen
  \bibfield  {author} {\bibinfo {author} {\bibfnamefont {P.~D.}\ \bibnamefont
  {Nation}}, \bibinfo {author} {\bibfnamefont {J.}~\bibnamefont {Suh}}, \ and\
  \bibinfo {author} {\bibfnamefont {M.~P.}\ \bibnamefont {Blencowe}},\ }\href
  {\doibase 10.1103/PhysRevA.93.022510} {\bibfield  {journal} {\bibinfo
  {journal} {Physical Review A}\ }\textbf {\bibinfo {volume} {93}},\ \bibinfo
  {pages} {022510} (\bibinfo {year} {2016})}\BibitemShut {NoStop}%
\bibitem [{\citenamefont {Neumeier}\ \emph {et~al.}(2018)\citenamefont
  {Neumeier}, \citenamefont {Northup},\ and\ \citenamefont
  {Chang}}]{bib:Neumeier2018a}%
  \BibitemOpen
  \bibfield  {author} {\bibinfo {author} {\bibfnamefont {L.}~\bibnamefont
  {Neumeier}}, \bibinfo {author} {\bibfnamefont {T.~E.}\ \bibnamefont
  {Northup}}, \ and\ \bibinfo {author} {\bibfnamefont {D.~E.}\ \bibnamefont
  {Chang}},\ }\href {\doibase 10.1103/PhysRevA.97.063857} {\bibfield  {journal}
  {\bibinfo  {journal} {Physical Review A}\ }\textbf {\bibinfo {volume} {97}},\
  \bibinfo {pages} {063857} (\bibinfo {year} {2018})}\BibitemShut {NoStop}%
\bibitem [{\citenamefont {Neumeier}\ and\ \citenamefont
  {Chang}(2018)}]{bib:Neumeier2018b}%
  \BibitemOpen
  \bibfield  {author} {\bibinfo {author} {\bibfnamefont {L.}~\bibnamefont
  {Neumeier}}\ and\ \bibinfo {author} {\bibfnamefont {D.~E.}\ \bibnamefont
  {Chang}},\ }\href {https://doi.org/10.1088/1367-2630/aad497} {\bibfield
  {journal} {\bibinfo  {journal} {New Journal of Physics}\ }\textbf {\bibinfo
  {volume} {20}},\ \bibinfo {pages} {083004} (\bibinfo {year}
  {2018})}\BibitemShut {NoStop}%
\bibitem [{\citenamefont {Settineri}\ \emph {et~al.}(2018)\citenamefont
  {Settineri}, \citenamefont {Macr\'{\i}}, \citenamefont {Ridolfo},
  \citenamefont {Di~Stefano}, \citenamefont {Kockum}, \citenamefont {Nori},\
  and\ \citenamefont {Savasta}}]{bib:Settineri2018}%
  \BibitemOpen
  \bibfield  {author} {\bibinfo {author} {\bibfnamefont {A.}~\bibnamefont
  {Settineri}}, \bibinfo {author} {\bibfnamefont {V.}~\bibnamefont
  {Macr\'{\i}}}, \bibinfo {author} {\bibfnamefont {A.}~\bibnamefont {Ridolfo}},
  \bibinfo {author} {\bibfnamefont {O.}~\bibnamefont {Di~Stefano}}, \bibinfo
  {author} {\bibfnamefont {A.~F.}\ \bibnamefont {Kockum}}, \bibinfo {author}
  {\bibfnamefont {F.}~\bibnamefont {Nori}}, \ and\ \bibinfo {author}
  {\bibfnamefont {S.}~\bibnamefont {Savasta}},\ }\href {\doibase
  10.1103/PhysRevA.98.053834} {\bibfield  {journal} {\bibinfo  {journal}
  {Physical Review A}\ }\textbf {\bibinfo {volume} {98}},\ \bibinfo {pages}
  {053834} (\bibinfo {year} {2018})}\BibitemShut {NoStop}%
\bibitem [{\citenamefont {Liao}\ \emph {et~al.}(2020)\citenamefont {Liao},
  \citenamefont {Huang}, \citenamefont {Tian}, \citenamefont {Kuang},\ and\
  \citenamefont {Sun}}]{bib:Liao2020}%
  \BibitemOpen
  \bibfield  {author} {\bibinfo {author} {\bibfnamefont {J.-Q.}\ \bibnamefont
  {Liao}}, \bibinfo {author} {\bibfnamefont {J.-F.}\ \bibnamefont {Huang}},
  \bibinfo {author} {\bibfnamefont {L.}~\bibnamefont {Tian}}, \bibinfo {author}
  {\bibfnamefont {L.-M.}\ \bibnamefont {Kuang}}, \ and\ \bibinfo {author}
  {\bibfnamefont {C.-P.}\ \bibnamefont {Sun}},\ }\href {\doibase
  10.1103/PhysRevA.101.063802} {\bibfield  {journal} {\bibinfo  {journal}
  {Physical Review A}\ }\textbf {\bibinfo {volume} {101}},\ \bibinfo {pages}
  {063802} (\bibinfo {year} {2020})}\BibitemShut {NoStop}%
\bibitem [{\citenamefont {Kounalakis}\ \emph {et~al.}(2020)\citenamefont
  {Kounalakis}, \citenamefont {Blanter},\ and\ \citenamefont
  {Steele}}]{bib:Kounalaksi2020}%
  \BibitemOpen
  \bibfield  {author} {\bibinfo {author} {\bibfnamefont {M.}~\bibnamefont
  {Kounalakis}}, \bibinfo {author} {\bibfnamefont {Y.~M.}\ \bibnamefont
  {Blanter}}, \ and\ \bibinfo {author} {\bibfnamefont {G.~A.}\ \bibnamefont
  {Steele}},\ }\href {\doibase 10.1103/PhysRevResearch.2.023335} {\bibfield
  {journal} {\bibinfo  {journal} {Physical Review Research}\ }\textbf {\bibinfo
  {volume} {2}},\ \bibinfo {pages} {023335} (\bibinfo {year}
  {2020})}\BibitemShut {NoStop}%
\bibitem [{\citenamefont {Ashhab}\ and\ \citenamefont
  {Nori}(2010)}]{bib:Ashhab2010}%
  \BibitemOpen
  \bibfield  {author} {\bibinfo {author} {\bibfnamefont {S.}~\bibnamefont
  {Ashhab}}\ and\ \bibinfo {author} {\bibfnamefont {F.}~\bibnamefont {Nori}},\
  }\href {\doibase 10.1103/PhysRevA.81.042311} {\bibfield  {journal} {\bibinfo
  {journal} {Physical Review A}\ }\textbf {\bibinfo {volume} {81}},\ \bibinfo
  {pages} {042311} (\bibinfo {year} {2010})}\BibitemShut {NoStop}%
\bibitem [{\citenamefont {Gu}\ \emph {et~al.}(2014)\citenamefont {Gu},
  \citenamefont {Li}, \citenamefont {Wu},\ and\ \citenamefont
  {Yang}}]{bib:Gu2014}%
  \BibitemOpen
  \bibfield  {author} {\bibinfo {author} {\bibfnamefont {W.-j.}\ \bibnamefont
  {Gu}}, \bibinfo {author} {\bibfnamefont {G.-x.}\ \bibnamefont {Li}}, \bibinfo
  {author} {\bibfnamefont {S.-p.}\ \bibnamefont {Wu}}, \ and\ \bibinfo {author}
  {\bibfnamefont {Y.-p.}\ \bibnamefont {Yang}},\ }\href
  {https://doi.org/10.1364/OE.22.018254} {\bibfield  {journal} {\bibinfo
  {journal} {Optics Express}\ }\textbf {\bibinfo {volume} {22}},\ \bibinfo
  {pages} {18254} (\bibinfo {year} {2014})}\BibitemShut {NoStop}%
\bibitem [{\citenamefont {S\'anchez Mu\~noz}\ \emph {et~al.}(2018)\citenamefont
  {S\'anchez Mu\~noz}, \citenamefont {Lara}, \citenamefont {Puebla},\ and\
  \citenamefont {Nori}}]{bib:Sanchez2018}%
  \BibitemOpen
  \bibfield  {author} {\bibinfo {author} {\bibfnamefont {C.}~\bibnamefont
  {S\'anchez Mu\~noz}}, \bibinfo {author} {\bibfnamefont {A.}~\bibnamefont
  {Lara}}, \bibinfo {author} {\bibfnamefont {J.}~\bibnamefont {Puebla}}, \ and\
  \bibinfo {author} {\bibfnamefont {F.}~\bibnamefont {Nori}},\ }\href {\doibase
  10.1103/PhysRevLett.121.123604} {\bibfield  {journal} {\bibinfo  {journal}
  {Physical Review Letters}\ }\textbf {\bibinfo {volume} {121}},\ \bibinfo
  {pages} {123604} (\bibinfo {year} {2018})}\BibitemShut {NoStop}%
\bibitem [{\citenamefont {Marinkovi\ifmmode~\acute{c}\else \'{c}\fi{}}\ \emph
  {et~al.}(2018)\citenamefont {Marinkovi\ifmmode~\acute{c}\else \'{c}\fi{}},
  \citenamefont {Wallucks}, \citenamefont {Riedinger}, \citenamefont {Hong},
  \citenamefont {Aspelmeyer},\ and\ \citenamefont
  {Gr\"oblacher}}]{bib:Marinkovic2018}%
  \BibitemOpen
  \bibfield  {author} {\bibinfo {author} {\bibfnamefont {I.}~\bibnamefont
  {Marinkovi\ifmmode~\acute{c}\else \'{c}\fi{}}}, \bibinfo {author}
  {\bibfnamefont {A.}~\bibnamefont {Wallucks}}, \bibinfo {author}
  {\bibfnamefont {R.}~\bibnamefont {Riedinger}}, \bibinfo {author}
  {\bibfnamefont {S.}~\bibnamefont {Hong}}, \bibinfo {author} {\bibfnamefont
  {M.}~\bibnamefont {Aspelmeyer}}, \ and\ \bibinfo {author} {\bibfnamefont
  {S.}~\bibnamefont {Gr\"oblacher}},\ }\href {\doibase
  10.1103/PhysRevLett.121.220404} {\bibfield  {journal} {\bibinfo  {journal}
  {Physical Review Letters}\ }\textbf {\bibinfo {volume} {121}},\ \bibinfo
  {pages} {220404} (\bibinfo {year} {2018})}\BibitemShut {NoStop}%
\bibitem [{\citenamefont {Ockeloen-Korppi}\ \emph {et~al.}(2018)\citenamefont
  {Ockeloen-Korppi}, \citenamefont {Damsk{\"{a}}gg}, \citenamefont
  {Pirkkalainen}, \citenamefont {Asjad}, \citenamefont {Clerk}, \citenamefont
  {Massel}, \citenamefont {J.},\ and\ \citenamefont
  {Sillanp{\"{a}}{\"{a}}}}]{bib:Ockeloen2018}%
  \BibitemOpen
  \bibfield  {author} {\bibinfo {author} {\bibfnamefont {C.}~\bibnamefont
  {Ockeloen-Korppi}}, \bibinfo {author} {\bibfnamefont {E.}~\bibnamefont
  {Damsk{\"{a}}gg}}, \bibinfo {author} {\bibfnamefont {J.-M.}\ \bibnamefont
  {Pirkkalainen}}, \bibinfo {author} {\bibfnamefont {M.}~\bibnamefont {Asjad}},
  \bibinfo {author} {\bibfnamefont {A.~A.}\ \bibnamefont {Clerk}}, \bibinfo
  {author} {\bibfnamefont {F.}~\bibnamefont {Massel}}, \bibinfo {author}
  {\bibfnamefont {W.~M.}\ \bibnamefont {J.}}, \ and\ \bibinfo {author}
  {\bibfnamefont {M.~A.}\ \bibnamefont {Sillanp{\"{a}}{\"{a}}}},\ }\href
  {https://doi.org/10.1038/s41586-018-0038-x} {\bibfield  {journal} {\bibinfo
  {journal} {Nature}\ }\textbf {\bibinfo {volume} {556}},\ \bibinfo {pages}
  {478–482} (\bibinfo {year} {2018})}\BibitemShut {NoStop}%
\bibitem [{\citenamefont {{H. Devoret}}\ \emph {et~al.}(2004)\citenamefont {{H.
  Devoret}}, \citenamefont {Wallraff},\ and\ \citenamefont
  {Martinis}}]{bib:H.Devoret2004}%
  \BibitemOpen
  \bibfield  {author} {\bibinfo {author} {\bibfnamefont {M.}~\bibnamefont {{H.
  Devoret}}}, \bibinfo {author} {\bibfnamefont {A.}~\bibnamefont {Wallraff}}, \
  and\ \bibinfo {author} {\bibfnamefont {J.~M.}\ \bibnamefont {Martinis}},\
  }\href@noop {} {\emph {\bibinfo {title} {{Superconducting Qubits: A Short
  Review}}}}\ (\bibinfo {year} {2004})\BibitemShut {NoStop}%
\bibitem [{\citenamefont {Abdi}\ \emph {et~al.}(2015)\citenamefont {Abdi},
  \citenamefont {Pernpeintner}, \citenamefont {Gross}, \citenamefont {Huebl},\
  and\ \citenamefont {Hartmann}}]{bib:Abdi2015}%
  \BibitemOpen
  \bibfield  {author} {\bibinfo {author} {\bibfnamefont {M.}~\bibnamefont
  {Abdi}}, \bibinfo {author} {\bibfnamefont {M.}~\bibnamefont {Pernpeintner}},
  \bibinfo {author} {\bibfnamefont {R.}~\bibnamefont {Gross}}, \bibinfo
  {author} {\bibfnamefont {H.}~\bibnamefont {Huebl}}, \ and\ \bibinfo {author}
  {\bibfnamefont {M.~J.}\ \bibnamefont {Hartmann}},\ }\href {\doibase
  10.1103/PhysRevLett.114.173602} {\bibfield  {journal} {\bibinfo  {journal}
  {Physical Review Letters}\ }\textbf {\bibinfo {volume} {114}},\ \bibinfo
  {pages} {173602} (\bibinfo {year} {2015})}\BibitemShut {NoStop}%
\bibitem [{\citenamefont {Shevchuk}\ \emph {et~al.}(2017)\citenamefont
  {Shevchuk}, \citenamefont {Steele},\ and\ \citenamefont
  {Blanter}}]{bib:Shevchuk2017}%
  \BibitemOpen
  \bibfield  {author} {\bibinfo {author} {\bibfnamefont {O.}~\bibnamefont
  {Shevchuk}}, \bibinfo {author} {\bibfnamefont {G.~A.}\ \bibnamefont
  {Steele}}, \ and\ \bibinfo {author} {\bibfnamefont {Y.~M.}\ \bibnamefont
  {Blanter}},\ }\href {\doibase 10.1103/PhysRevB.96.014508} {\bibfield
  {journal} {\bibinfo  {journal} {Physical Review B}\ }\textbf {\bibinfo
  {volume} {96}},\ \bibinfo {pages} {014508} (\bibinfo {year}
  {2017})}\BibitemShut {NoStop}%
\bibitem [{\citenamefont {Blien}\ \emph {et~al.}(2020)\citenamefont {Blien},
  \citenamefont {Steger}, \citenamefont {H{\"{u}}ttner}, \citenamefont
  {Graaf},\ and\ \citenamefont {H{\"{u}}ttel}}]{bib:Blien2020}%
  \BibitemOpen
  \bibfield  {author} {\bibinfo {author} {\bibfnamefont {S.}~\bibnamefont
  {Blien}}, \bibinfo {author} {\bibfnamefont {P.}~\bibnamefont {Steger}},
  \bibinfo {author} {\bibfnamefont {N.}~\bibnamefont {H{\"{u}}ttner}}, \bibinfo
  {author} {\bibfnamefont {R.}~\bibnamefont {Graaf}}, \ and\ \bibinfo {author}
  {\bibfnamefont {A.~K.}\ \bibnamefont {H{\"{u}}ttel}},\ }\href {\doibase
  10.1038/s41467-020-15433-3} {\bibfield  {journal} {\bibinfo  {journal}
  {Nature Communications}\ }\textbf {\bibinfo {volume} {11}},\ \bibinfo {pages}
  {1636} (\bibinfo {year} {2020})}\BibitemShut {NoStop}%
\bibitem [{\citenamefont {Heikkil\"a}\ \emph {et~al.}(2014)\citenamefont
  {Heikkil\"a}, \citenamefont {Massel}, \citenamefont {Tuorila}, \citenamefont
  {Khan},\ and\ \citenamefont {Sillanp\"a\"a}}]{bib:Heikkila2014}%
  \BibitemOpen
  \bibfield  {author} {\bibinfo {author} {\bibfnamefont {T.~T.}\ \bibnamefont
  {Heikkil\"a}}, \bibinfo {author} {\bibfnamefont {F.}~\bibnamefont {Massel}},
  \bibinfo {author} {\bibfnamefont {J.}~\bibnamefont {Tuorila}}, \bibinfo
  {author} {\bibfnamefont {R.}~\bibnamefont {Khan}}, \ and\ \bibinfo {author}
  {\bibfnamefont {M.~A.}\ \bibnamefont {Sillanp\"a\"a}},\ }\href {\doibase
  10.1103/PhysRevLett.112.203603} {\bibfield  {journal} {\bibinfo  {journal}
  {Phys. Rev. Lett.}\ }\textbf {\bibinfo {volume} {112}},\ \bibinfo {pages}
  {203603} (\bibinfo {year} {2014})}\BibitemShut {NoStop}%
\bibitem [{\citenamefont {Rimberg}\ \emph {et~al.}(2014)\citenamefont
  {Rimberg}, \citenamefont {Blencowe}, \citenamefont {Armour},\ and\
  \citenamefont {Nation}}]{bib:Rimberg2014}%
  \BibitemOpen
  \bibfield  {author} {\bibinfo {author} {\bibfnamefont {A.~J.}\ \bibnamefont
  {Rimberg}}, \bibinfo {author} {\bibfnamefont {M.~P.}\ \bibnamefont
  {Blencowe}}, \bibinfo {author} {\bibfnamefont {A.~D.}\ \bibnamefont
  {Armour}}, \ and\ \bibinfo {author} {\bibfnamefont {P.~D.}\ \bibnamefont
  {Nation}},\ }\href {\doibase 10.1088/1367-2630/16/5/055008} {\bibfield
  {journal} {\bibinfo  {journal} {New Journal of Physics}\ }\textbf {\bibinfo
  {volume} {16}},\ \bibinfo {pages} {055008} (\bibinfo {year}
  {2014})}\BibitemShut {NoStop}%
\bibitem [{\citenamefont {Pirkkalainen}\ \emph {et~al.}(2015)\citenamefont
  {Pirkkalainen}, \citenamefont {Cho}, \citenamefont {Massel}, \citenamefont
  {Tuorila}, \citenamefont {Heikkil{\"{a}}}, \citenamefont {Hakonen},\ and\
  \citenamefont {Sillanp{\"{a}}{\"{a}}}}]{bib:Pirkkalainen2015}%
  \BibitemOpen
  \bibfield  {author} {\bibinfo {author} {\bibfnamefont {J.-M.}\ \bibnamefont
  {Pirkkalainen}}, \bibinfo {author} {\bibfnamefont {S.~U.}\ \bibnamefont
  {Cho}}, \bibinfo {author} {\bibfnamefont {F.}~\bibnamefont {Massel}},
  \bibinfo {author} {\bibfnamefont {J.}~\bibnamefont {Tuorila}}, \bibinfo
  {author} {\bibfnamefont {T.~T.}\ \bibnamefont {Heikkil{\"{a}}}}, \bibinfo
  {author} {\bibfnamefont {P.~J.}\ \bibnamefont {Hakonen}}, \ and\ \bibinfo
  {author} {\bibfnamefont {M.~A.}\ \bibnamefont {Sillanp{\"{a}}{\"{a}}}},\
  }\href {http://dx.doi.org/10.1038/ncomms7981} {\bibfield  {journal} {\bibinfo
   {journal} {Nature Communications}\ }\textbf {\bibinfo {volume} {6}},\
  \bibinfo {pages} {6981} (\bibinfo {year} {2015})}\BibitemShut {NoStop}%
\bibitem [{\citenamefont {Averin}\ \emph {et~al.}(1985)\citenamefont {Averin},
  \citenamefont {Zorin},\ and\ \citenamefont {Likharev}}]{bib:Averin1985}%
  \BibitemOpen
  \bibfield  {author} {\bibinfo {author} {\bibfnamefont {D.~V.}\ \bibnamefont
  {Averin}}, \bibinfo {author} {\bibfnamefont {A.~B.}\ \bibnamefont {Zorin}}, \
  and\ \bibinfo {author} {\bibfnamefont {K.~K.}\ \bibnamefont {Likharev}},\
  }\href {http://jetp.ac.ru/cgi-bin/dn/e_061_02_0407.pdf} {\bibfield  {journal}
  {\bibinfo  {journal} {Sov. Phys. JETP}\ }\textbf {\bibinfo {volume} {61}},\
  \bibinfo {pages} {407} (\bibinfo {year} {1985})}\BibitemShut {NoStop}%
\bibitem [{\citenamefont {Averin}\ and\ \citenamefont
  {Bruder}(2003)}]{bib:Averin2003}%
  \BibitemOpen
  \bibfield  {author} {\bibinfo {author} {\bibfnamefont {D.~V.}\ \bibnamefont
  {Averin}}\ and\ \bibinfo {author} {\bibfnamefont {C.}~\bibnamefont
  {Bruder}},\ }\href {\doibase 10.1103/PhysRevLett.91.057003} {\bibfield
  {journal} {\bibinfo  {journal} {Physical Review Letters}\ }\textbf {\bibinfo
  {volume} {91}},\ \bibinfo {pages} {057003} (\bibinfo {year}
  {2003})}\BibitemShut {NoStop}%
\bibitem [{\citenamefont {Sillanp{\"{a}}{\"{a}}}\ \emph
  {et~al.}(2005)\citenamefont {Sillanp{\"{a}}{\"{a}}}, \citenamefont
  {Lehtinen}, \citenamefont {Paila}, \citenamefont {Makhlin}, \citenamefont
  {Roschier},\ and\ \citenamefont {Hakonen}}]{bib:Sillanpaa2005}%
  \BibitemOpen
  \bibfield  {author} {\bibinfo {author} {\bibfnamefont {M.~A.}\ \bibnamefont
  {Sillanp{\"{a}}{\"{a}}}}, \bibinfo {author} {\bibfnamefont {T.}~\bibnamefont
  {Lehtinen}}, \bibinfo {author} {\bibfnamefont {A.}~\bibnamefont {Paila}},
  \bibinfo {author} {\bibfnamefont {Y.}~\bibnamefont {Makhlin}}, \bibinfo
  {author} {\bibfnamefont {L.}~\bibnamefont {Roschier}}, \ and\ \bibinfo
  {author} {\bibfnamefont {P.~J.}\ \bibnamefont {Hakonen}},\ }\href {\doibase
  10.1103/PhysRevLett.95.206806} {\bibfield  {journal} {\bibinfo  {journal}
  {Physical Review Letters}\ }\textbf {\bibinfo {volume} {95}},\ \bibinfo
  {pages} {206806} (\bibinfo {year} {2005})}\BibitemShut {NoStop}%
\bibitem [{\citenamefont {Duty}\ \emph {et~al.}(2005)\citenamefont {Duty},
  \citenamefont {Johansson}, \citenamefont {Bladh}, \citenamefont {Gunnarsson},
  \citenamefont {Wilson},\ and\ \citenamefont {Delsing}}]{bib:Duty2005}%
  \BibitemOpen
  \bibfield  {author} {\bibinfo {author} {\bibfnamefont {T.}~\bibnamefont
  {Duty}}, \bibinfo {author} {\bibfnamefont {G.}~\bibnamefont {Johansson}},
  \bibinfo {author} {\bibfnamefont {K.}~\bibnamefont {Bladh}}, \bibinfo
  {author} {\bibfnamefont {D.}~\bibnamefont {Gunnarsson}}, \bibinfo {author}
  {\bibfnamefont {C.}~\bibnamefont {Wilson}}, \ and\ \bibinfo {author}
  {\bibfnamefont {P.}~\bibnamefont {Delsing}},\ }\href {\doibase
  10.1103/PhysRevLett.95.206807} {\bibfield  {journal} {\bibinfo  {journal}
  {Physical Review Letters}\ }\textbf {\bibinfo {volume} {95}},\ \bibinfo
  {pages} {206807} (\bibinfo {year} {2005})}\BibitemShut {NoStop}%
\bibitem [{\citenamefont {Persson}\ \emph {et~al.}(2010)\citenamefont
  {Persson}, \citenamefont {Wilson}, \citenamefont {Sandberg},\ and\
  \citenamefont {Delsing}}]{bib:Persson2010}%
  \BibitemOpen
  \bibfield  {author} {\bibinfo {author} {\bibfnamefont {F.}~\bibnamefont
  {Persson}}, \bibinfo {author} {\bibfnamefont {C.~M.}\ \bibnamefont {Wilson}},
  \bibinfo {author} {\bibfnamefont {M.}~\bibnamefont {Sandberg}}, \ and\
  \bibinfo {author} {\bibfnamefont {P.}~\bibnamefont {Delsing}},\ }\href
  {\doibase 10.1103/PhysRevB.82.134533} {\bibfield  {journal} {\bibinfo
  {journal} {Physical Review B}\ }\textbf {\bibinfo {volume} {82}},\ \bibinfo
  {pages} {134533} (\bibinfo {year} {2010})}\BibitemShut {NoStop}%
\bibitem [{\citenamefont {Roschier}\ \emph {et~al.}(2005)\citenamefont
  {Roschier}, \citenamefont {Sillanp\"a\"a},\ and\ \citenamefont
  {Hakonen}}]{bib:Roschier2005}%
  \BibitemOpen
  \bibfield  {author} {\bibinfo {author} {\bibfnamefont {L.}~\bibnamefont
  {Roschier}}, \bibinfo {author} {\bibfnamefont {M.}~\bibnamefont
  {Sillanp\"a\"a}}, \ and\ \bibinfo {author} {\bibfnamefont {P.}~\bibnamefont
  {Hakonen}},\ }\href {\doibase 10.1103/PhysRevB.71.024530} {\bibfield
  {journal} {\bibinfo  {journal} {Physical Review B}\ }\textbf {\bibinfo
  {volume} {71}},\ \bibinfo {pages} {024530} (\bibinfo {year}
  {2005})}\BibitemShut {NoStop}%
\bibitem [{\citenamefont {Shaw}\ \emph {et~al.}(2009)\citenamefont {Shaw},
  \citenamefont {Bueno}, \citenamefont {Day}, \citenamefont {Bradford},\ and\
  \citenamefont {Echternach}}]{bib:Shaw2009}%
  \BibitemOpen
  \bibfield  {author} {\bibinfo {author} {\bibfnamefont {M.~D.}\ \bibnamefont
  {Shaw}}, \bibinfo {author} {\bibfnamefont {J.}~\bibnamefont {Bueno}},
  \bibinfo {author} {\bibfnamefont {P.}~\bibnamefont {Day}}, \bibinfo {author}
  {\bibfnamefont {C.~M.}\ \bibnamefont {Bradford}}, \ and\ \bibinfo {author}
  {\bibfnamefont {P.~M.}\ \bibnamefont {Echternach}},\ }\href {\doibase
  10.1103/PhysRevB.79.144511} {\bibfield  {journal} {\bibinfo  {journal}
  {Physical Review B}\ }\textbf {\bibinfo {volume} {79}},\ \bibinfo {pages}
  {144511} (\bibinfo {year} {2009})}\BibitemShut {NoStop}%
\bibitem [{\citenamefont {Imoto}\ \emph {et~al.}(1985)\citenamefont {Imoto},
  \citenamefont {Haus},\ and\ \citenamefont {Yamamoto}}]{bib:Imoto1985}%
  \BibitemOpen
  \bibfield  {author} {\bibinfo {author} {\bibfnamefont {N.}~\bibnamefont
  {Imoto}}, \bibinfo {author} {\bibfnamefont {H.~A.}\ \bibnamefont {Haus}}, \
  and\ \bibinfo {author} {\bibfnamefont {Y.}~\bibnamefont {Yamamoto}},\ }\href
  {\doibase 10.1103/PhysRevA.32.2287} {\bibfield  {journal} {\bibinfo
  {journal} {Physical Review A}\ }\textbf {\bibinfo {volume} {32}},\ \bibinfo
  {pages} {2287} (\bibinfo {year} {1985})}\BibitemShut {NoStop}%
\bibitem [{\citenamefont {Khan}\ \emph {et~al.}(2015)\citenamefont {Khan},
  \citenamefont {Massel},\ and\ \citenamefont {Heikkil\"a}}]{bib:Khan2015}%
  \BibitemOpen
  \bibfield  {author} {\bibinfo {author} {\bibfnamefont {R.}~\bibnamefont
  {Khan}}, \bibinfo {author} {\bibfnamefont {F.}~\bibnamefont {Massel}}, \ and\
  \bibinfo {author} {\bibfnamefont {T.~T.}\ \bibnamefont {Heikkil\"a}},\ }\href
  {\doibase 10.1103/PhysRevA.91.043822} {\bibfield  {journal} {\bibinfo
  {journal} {Physical Review A}\ }\textbf {\bibinfo {volume} {91}},\ \bibinfo
  {pages} {043822} (\bibinfo {year} {2015})}\BibitemShut {NoStop}%
\bibitem [{\citenamefont {Sarala}\ and\ \citenamefont
  {Massel}(2015)}]{bib:Sarala2015}%
  \BibitemOpen
  \bibfield  {author} {\bibinfo {author} {\bibfnamefont {R.}~\bibnamefont
  {Sarala}}\ and\ \bibinfo {author} {\bibfnamefont {F.}~\bibnamefont
  {Massel}},\ }\href {http://arxiv.org/abs/1509.00964} {\bibfield  {journal}
  {\bibinfo  {journal} {Nanoscale Systems}\ }\textbf {\bibinfo {volume} {4}},\
  \bibinfo {pages} {18} (\bibinfo {year} {2015})}\BibitemShut {NoStop}%
\bibitem [{\citenamefont {Xiong}\ \emph {et~al.}(2016)\citenamefont {Xiong},
  \citenamefont {Jin}, \citenamefont {Qiu}, \citenamefont {Lam},\ and\
  \citenamefont {You}}]{bib:Xiong2016}%
  \BibitemOpen
  \bibfield  {author} {\bibinfo {author} {\bibfnamefont {W.}~\bibnamefont
  {Xiong}}, \bibinfo {author} {\bibfnamefont {D.-Y.}\ \bibnamefont {Jin}},
  \bibinfo {author} {\bibfnamefont {Y.}~\bibnamefont {Qiu}}, \bibinfo {author}
  {\bibfnamefont {C.-H.}\ \bibnamefont {Lam}}, \ and\ \bibinfo {author}
  {\bibfnamefont {J.~Q.}\ \bibnamefont {You}},\ }\href {\doibase
  10.1103/PhysRevA.93.023844} {\bibfield  {journal} {\bibinfo  {journal}
  {Physical Review A}\ }\textbf {\bibinfo {volume} {93}},\ \bibinfo {pages}
  {023844} (\bibinfo {year} {2016})}\BibitemShut {NoStop}%
\bibitem [{\citenamefont {Chakraborty}\ and\ \citenamefont
  {Sarma}(2017)}]{bib:Chakraborty2017}%
  \BibitemOpen
  \bibfield  {author} {\bibinfo {author} {\bibfnamefont {S.}~\bibnamefont
  {Chakraborty}}\ and\ \bibinfo {author} {\bibfnamefont {A.~K.}\ \bibnamefont
  {Sarma}},\ }\href {\doibase 10.1364/JOSAB.34.001503} {\bibfield  {journal}
  {\bibinfo  {journal} {J. Opt. Soc. Am. B}\ }\textbf {\bibinfo {volume}
  {34}},\ \bibinfo {pages} {1503} (\bibinfo {year} {2017})}\BibitemShut
  {NoStop}%
\bibitem [{\citenamefont {Yin}\ \emph {et~al.}(2018)\citenamefont {Yin},
  \citenamefont {L\"{u}}, \citenamefont {Wan}, \citenamefont {Bin},\ and\
  \citenamefont {Wu}}]{bib:Yin2018}%
  \BibitemOpen
  \bibfield  {author} {\bibinfo {author} {\bibfnamefont {T.-S.}\ \bibnamefont
  {Yin}}, \bibinfo {author} {\bibfnamefont {X.-Y.}\ \bibnamefont {L\"{u}}},
  \bibinfo {author} {\bibfnamefont {L.-L.}\ \bibnamefont {Wan}}, \bibinfo
  {author} {\bibfnamefont {S.-W.}\ \bibnamefont {Bin}}, \ and\ \bibinfo
  {author} {\bibfnamefont {Y.}~\bibnamefont {Wu}},\ }\href {\doibase
  10.1364/OL.43.002050} {\bibfield  {journal} {\bibinfo  {journal} {Optics
  Letters}\ }\textbf {\bibinfo {volume} {43}},\ \bibinfo {pages} {2050}
  (\bibinfo {year} {2018})}\BibitemShut {NoStop}%
\bibitem [{\citenamefont {Kounalakis}\ \emph {et~al.}(2018)\citenamefont
  {Kounalakis}, \citenamefont {Dickel}, \citenamefont {Bruno}, \citenamefont
  {Langford},\ and\ \citenamefont {Steele}}]{bib:Kounalakis2018}%
  \BibitemOpen
  \bibfield  {author} {\bibinfo {author} {\bibfnamefont {M.}~\bibnamefont
  {Kounalakis}}, \bibinfo {author} {\bibfnamefont {C.}~\bibnamefont {Dickel}},
  \bibinfo {author} {\bibfnamefont {A.}~\bibnamefont {Bruno}}, \bibinfo
  {author} {\bibfnamefont {N.~K.}\ \bibnamefont {Langford}}, \ and\ \bibinfo
  {author} {\bibfnamefont {G.~A.}\ \bibnamefont {Steele}},\ }\href
  {https://doi.org/10.1038/s41534-018-0088-9} {\bibfield  {journal} {\bibinfo
  {journal} {npj Quantum Information}\ }\textbf {\bibinfo {volume} {4}},\
  \bibinfo {pages} {38} (\bibinfo {year} {2018})}\BibitemShut {NoStop}%
\bibitem [{\citenamefont {Joyez}\ \emph {et~al.}(1994)\citenamefont {Joyez},
  \citenamefont {Lafarge}, \citenamefont {Filipe}, \citenamefont {Esteve},\
  and\ \citenamefont {Devoret}}]{bib:Joyez1994}%
  \BibitemOpen
  \bibfield  {author} {\bibinfo {author} {\bibfnamefont {P.}~\bibnamefont
  {Joyez}}, \bibinfo {author} {\bibfnamefont {P.}~\bibnamefont {Lafarge}},
  \bibinfo {author} {\bibfnamefont {A.}~\bibnamefont {Filipe}}, \bibinfo
  {author} {\bibfnamefont {D.}~\bibnamefont {Esteve}}, \ and\ \bibinfo {author}
  {\bibfnamefont {M.~H.}\ \bibnamefont {Devoret}},\ }\href {\doibase
  10.1103/PhysRevLett.72.2458} {\bibfield  {journal} {\bibinfo  {journal}
  {Physical Review Letters}\ }\textbf {\bibinfo {volume} {72}},\ \bibinfo
  {pages} {2458} (\bibinfo {year} {1994})}\BibitemShut {NoStop}%
\bibitem [{\citenamefont {Aumentado}\ \emph {et~al.}(2004)\citenamefont
  {Aumentado}, \citenamefont {Keller}, \citenamefont {Martinis},\ and\
  \citenamefont {Devoret}}]{bib:Aumentado2004}%
  \BibitemOpen
  \bibfield  {author} {\bibinfo {author} {\bibfnamefont {J.}~\bibnamefont
  {Aumentado}}, \bibinfo {author} {\bibfnamefont {M.~W.}\ \bibnamefont
  {Keller}}, \bibinfo {author} {\bibfnamefont {J.~M.}\ \bibnamefont
  {Martinis}}, \ and\ \bibinfo {author} {\bibfnamefont {M.~H.}\ \bibnamefont
  {Devoret}},\ }\href {\doibase 10.1103/PhysRevLett.92.066802} {\bibfield
  {journal} {\bibinfo  {journal} {Physical Review Letters}\ }\textbf {\bibinfo
  {volume} {92}},\ \bibinfo {pages} {066802} (\bibinfo {year}
  {2004})}\BibitemShut {NoStop}%
\bibitem [{\citenamefont {Vool}\ and\ \citenamefont
  {Devoret}(2017)}]{bib:Vool2017}%
  \BibitemOpen
  \bibfield  {author} {\bibinfo {author} {\bibfnamefont {U.}~\bibnamefont
  {Vool}}\ and\ \bibinfo {author} {\bibfnamefont {M.~H.}\ \bibnamefont
  {Devoret}},\ }\href {\doibase 10.1002/cta.2359} {\bibfield  {journal}
  {\bibinfo  {journal} {Int. J. Circ. Theor. Appl.}\ }\textbf {\bibinfo
  {volume} {45}},\ \bibinfo {pages} {897} (\bibinfo {year} {2017})}\BibitemShut
  {NoStop}%
\end{thebibliography}%

\end{document}